\documentclass[english,a4paper,11pt]{article}
\usepackage[margin=3cm]{geometry}
\usepackage[latin1]{inputenc}
\usepackage{babel}
\usepackage{bm}
\usepackage{amsmath,amsthm}
\usepackage{latexsym}
\usepackage{booktabs}
\usepackage[final]{graphicx}
\DeclareGraphicsExtensions{.jpg,.jpeg,.pdf,.png,.mps}
\usepackage{epsfig}
\usepackage[round]{natbib}
\usepackage{natbib}
\setcitestyle{aysep={}} 
\usepackage{rotating}
\usepackage{color}
\usepackage{xcolor}
\usepackage{colortbl}
\usepackage[bitstream-charter]{mathdesign}
\usepackage[T1]{fontenc}
\usepackage{threeparttable}
\usepackage{float}
\usepackage{url} 
\usepackage{appendix}
\usepackage{hyperref}
\usepackage{algorithm}
\usepackage{pdflscape}
\usepackage[misc]{ifsym}
\hypersetup{
     colorlinks   = true,
     citecolor    = blue
}

\usepackage{silence}
\newcommand{\vir}[1]{``#1''}

\title{Constrained mixtures of generalized normal distributions}

\author{$\mathrm{Pierdomenico \ Duttilo}^\mathrm{*,\hspace{0.5mm}\textrm{\Letter}},  
	\  \mathrm{Stefano \ Antonio\ Gattone}^\mathrm{2},
	\  \mathrm{Alfred \ Kume}^\mathrm{3}$\\  
	$^\mathrm{1}$\small{\emph{Department of Statistical Sciences, University of Padova, Padova, Italy}}\\
    $^\mathrm{2}$\small{\emph{DiSEGS, University "G. d'Annunzio" of Chieti-Pescara, Pescara, Italy}}\\
    $^\mathrm{3}$\small{\emph{School of Mathematics, Statistics and Actuarial Sciences, University of Kent, Canterbury, UK}}\\
    $^\mathrm{\textrm{\Letter}}$\small{\emph{Corresponding author: pierdomenico.duttilo@unipd.it}}
}
\date{}

\begin{document}
	\maketitle
\begin{abstract}
This work introduces a family of univariate constrained mixtures of generalized normal distributions (CMGND) where the location, scale, and shape parameters can be constrained to be equal across any subset of mixture components. An expectation conditional maximisation (ECM) algorithm with Newton-Raphson updates is used to estimate the model parameters under the constraints. Simulation studies demonstrate that imposing correct constraints leads to more accurate parameter estimation compared to unconstrained mixtures, especially when components substantially overlap. Constrained models also exhibit competitive performance in capturing key characteristics of the marginal distribution, such as kurtosis. On a real dataset of daily stock index returns, CMGND models outperform constrained mixtures of normals and Student's t distributions based on the BIC criterion, highlighting their flexibility in modelling nonnormal features. The proposed constrained approach enhances interpretability and can improve parametric efficiency without compromising distributional flexibility for complex data.

\noindent 
\hspace{1cm} 
\end{abstract}

\section{Introduction}\label{sec1}
Finite mixtures of distributions are widely used to analyse complex distributions of data \citep{McLachlan2019} since \cite{Pearson1984} first introduced a mixture of two normal distributions. Normal mixture modelling is a well-known method that is used for most applications in a wide range of fields. However, datasets characterised by nonnormal features, such as asymmetry, multimodality, leptokurtosis, and heavy tails, require more flexible tools. Hence, \vir{\textit{non-normal model-based methods}} \citep{McLachlan2013} have attracted the attention of researchers. 

As argued by \cite{Nguyen2014} mixtures of generalized normal distributions (MGND) \vir{\textit{have the flexibility required to fit the shape of the data better than the Gaussian mixture model}}. The generalized normal distribution (GND) is \textit{\vir{natural generalisation of the normal distribution}} \citep{Nadarajah2005} and able to model a wide variety of statistical behaviours thanks to the additional shape parameter that controls the tail behaviour. MGND have been successfully applied in computer vision and pattern recognition problems. For example, \cite{Bazi2006} and \cite{Allili2008} applied univariate MGND for image processing. Parameter estimation was performed using the maximum likelihood estimation (MLE) and the expectation-maximisation (EM) algorithm. The numerical optimisation based on the Newton-Raphson
method is used, since the system to resolve the updating equation of the shape parameter is heavily nonlinear. As an alternative, \cite{Mohamed2009} estimated the shape parameter using the analytical relationship between the shape parameter and kurtosis. \cite{Nguyen2014} proposed a generalised univariate Gaussian mixture model defining a bounded support region in $\mathbb{R}$ for each component.

Recently, \cite{Wen2020} studied a two-component MGND model and proposed an expectation conditional maximisation (ECM) algorithm for parameter estimation. In particular, they show that for the modelling purposes of the S\&P 500 and Shanghai Stock Exchange Composite Index (SSEC), such mixtures outperform those constructed by simply using mixtures of normals. Finally, \cite{mgnd2025} showed that the estimation of the shape parameter within the maximum likelihood estimation framework can lead to numerical and degeneracy issues, especially when the shape parameter is greater than 2. They introduced an expectation conditional maximisation algorithm with step size (ECMs) incorporated with two key innovations: an adaptive step size function for the Newton-Raphson update of the shape parameter and a modified stopping criterion for the EM iterations where the shape parameter update is skipped if the first derivative of the Q-function with respect to the shape parameter falls below a given threshold.

It is well known that if the parameters of some mixture model are not properly constrained, the resulting likelihood is unbounded \cite{mclachlan2000}. For this reason, the maximum likelihood estimation of Gaussian mixture models is problematic for both univariate and multivariate data. For the univariate case, the problem occurs in the presence of a fitted component that has a very small estimate of the variance, that is, a few data points relatively close together \citep{Biernacki2003}. As a result, the likelihood function either increases intolerably high iteratively (degeneracy) or leads to spurious solutions, i.e., with a high likelihood and also a high bias. Existing methods to overcome this drawback are based on the seminal work of \cite{Hathaway1986}, who, in the univariate case, imposed a lower bound on the ratios of the scale parameters. Similarly, in the multivariate case, the lower bound is imposed on the eigenvalues of each pair of covariance matrices \citep{Banfield1993,Celeux1995,Roccietal2018}.

In order to handle the degeneracy of the likelihood function, in the pathological context of univariate normal mixtures, \cite{Quandt1978} and \cite{Chauveau2013} consider imposing linear constraints on the mean and variance parameters. Similarly, \cite{Andrews2011,Andrews2018,Massing2021} presented a family of univariate mixtures of Student t components, while requiring that variances and degrees of freedom remain equal across the mixture components. 

The use of constraints clearly leads to more parsimonious models as the number of unconstrained parameters increases linearly with the (potentially high) number of components. More specifically for the univariate MGND, the unknown number of parameters is $4K-1$ where $K$ is the number of components of the mixture. For large $K$, this can be an issue, while in some circumstances the nature of the constraints could also be related to the underlying data domain.
 
To our knowledge, none of the existing studies implements equality constraints on the parameters of the univariate MGND. We therefore propose and explore such constrained mixtures for the generalized normal distributions (CMGND) where the parameters are allowed to be equal across any subset of mixture components. Since GND is a generalisation of the normal distribution, it is straightforward to observe that the aforementioned degeneracy in MLE for normal mixtures also holds for MGND. In the specific context of MGND, where the estimation of the shape parameter is complex and can lead to numerical and degeneracy problems \citep{mgnd2025, Deledalle18, Roenko2014}, forcing some shape parameters to be equal between different components of the mixture could help stabilise the estimation process and prevent spurious solutions. Furthermore, the implementation of such constraints not only solves the numerical degeneracy of the log-likelihood but also enhances the interpretability and parametric efficiency of the final solution.

We consider constraints imposed on the location and/or scale and/or shape parameters. The MLE of the parameters is obtained via the ECM algorithm \citep{mgnd2025}. To implement the constraints between the different components, we need to modify the conditional maximisation phase of the algorithm with the derivation of new update equations that reflect the relationship between the parameters of the different components. The \texttt{R} package \texttt{cmgnd} is available to estimate the proposed constrained MGND models \citep{DuttiloR2024}.

The proposed methodology is tested on simulated data across different scenarios such as models with common scales and / or common kurtosis (that is, mixtures with a common shape parameter). Furthermore,a comparative analysis on the 50 constituents of the Euro Stoxx 50 index is performed. The goodness-of-fit of the CMGND model is compared to that of the constrained mixture of normals and the constrained mixture of Student-t distributions. The results clearly show that CMGND models could be advantageous in certain situations where the reduction in complexity does not compromise the model's ability to describe the data. 

The remainder of the paper is organised as follows. The proposed constrained mixtures of generalized normal distributions are presented in Section \ref{CMGNDmeth} where we detail the MLE method of the parameters using the ECM algorithm. The performance of the CMGND is explored in simulated and real data sets in Sections \ref{SM} and \ref{RDA}, respectively. We conclude with some general remarks in Section \ref{Conclusions}.

\section{Constrained mixtures of generalized normal distributions}\label{CMGNDmeth} 
\begin{figure}[H]
\centering
\includegraphics[width=12cm]{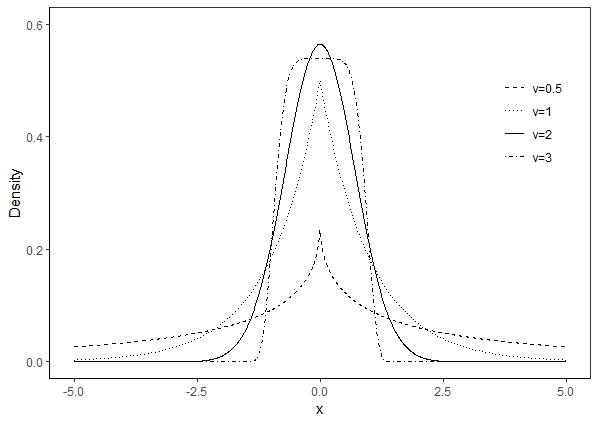}
\caption{The GND densities for $\mu=1$, $\sigma=1$ and different shape values.}\label{DensityGND}	
\end{figure}

A random variable $X$ is said to have the GND with parameters $\mu$ (location), $\sigma$ (scale) and $\nu$ (shape) if its probability density function (pdf) is given by
\begin{equation}
f(x|\mu,\sigma,\nu)=\frac{\nu}{2\sigma\Gamma(1\mathbin{/}\nu)}\exp\Biggr\{-\Biggr|\frac{x-\mu}{\sigma}\Biggr|^\nu\Biggr\},
\label{GND}
\end{equation}
with $\Gamma(1\mathbin{/}\nu)=\int_0^\infty t^{1\mathbin{/}\nu-1}\exp\{-t\}dt$, $-\infty<x<\infty$, $-\infty<\mu<\infty$, $\sigma>0$, $\nu>0$. Figure \ref{DensityGND} shows the probability density function of the GND (Eq. \ref{GND}) for $\mu=1$, $\sigma=1$ and different shape values. If $\nu=1$ the GND reduces to the Laplace distribution and if $\nu=2$ it coincides with the normal distribution. It is noticed that $1<\nu<2$ yields an \vir{intermediate distribution} between the
normal and the Laplace distribution. As limit cases, for $\nu\rightarrow\infty$ the distribution tends to a uniform
distribution, while for $\nu\rightarrow0$ it will be impulsive \citep{Nadarajah2005,Bazi2006,Dytso2018}.

A finite mixture of GND with K components is given by the marginal
distribution of the random variable $X$
\begin{equation}
F(x|\theta)=\sum_{k=1}^K\pi_kf_k(x|\mu_k, \sigma_k, \nu_k)\\
\label{MGND}
\end{equation}
where $f_k$ is defined as in Eq.(\ref{GND}), and the set of all mixture parameters is given by $\theta=\{\pi_k,\mu_k,\sigma_k,$ $\nu_k,k=1,...,K\}$ belonging to the parameter space $\Theta=\{\theta: 0<\pi_k<1, \sum_{k=1}^K\pi_k=1, \mu_k \in \text{R}, \sigma_k>0, \nu_k>0, k=1,...,K\}$ with $\text{dim}(\theta)=p$. Let define a partition of the set $\{1,2,\dots,K\}$ as $C_1,C_2,\dots,C_R$ such that $C_1 \cup C_2 \cup \dots \cup C_R=\{1,2,\dots,K\}$ and $C_i \cap C_j = \emptyset$  $\forall_{i,j}=1,2,\dots, K$. The parameter space $\Theta$ can be constrained by imposing equality constraints such as  $\mu_k=\mu_r$ and/or $\sigma_k=\sigma_r$ and/or $\nu_k=\nu_r$ for all $k \in C_r$. For $K=2$, Table \ref{8models} shows the 6-model family of CMGND together with the MGND model. By constraining the parameter space, the number of parameters ($p$) of each model is gradually reduced as shown in the last column of Table \ref{8models}. 

\begin{table}[H]
\caption{6-model family of CMGND and MGND models when $K=2$.}
\label{8models}
\centering
\resizebox{7cm}{!}{
	\begin{tabular}{lcccc}
		\toprule
\multicolumn{1}{c}{Model}&\multicolumn{1}{c}{\textit{$\mu_k$}}&\multicolumn{1}{c}{$\sigma_k$}&\multicolumn{1}{c}{$\nu_k$}&\multicolumn{1}{c}{$p$}
\\			
		\midrule
		MGND      & U & U & U & $4K-1$\\
		CMGND-CUU & C & U & U & $3K$ \\
        CMGND-UCU & U & C & U & $3K$ \\
        CMGND-UUC & U & U & C & $3K$ \\
        CMGND-CCU & C & C & U & $2K+1$ \\
        CMGND-CUC & C & U & C & $2K+1$ \\
        CMGND-UCC & U & C & C & $2K+1$ \\
		\bottomrule
		\end{tabular}
        }
    \begin{tablenotes}
    \centering
    \item[]{\footnotesize Note. C denotes constrained, U denotes unconstrained. $K$ is the number of mixture components.}
    \end{tablenotes}
\end{table}

\subsection{Parameters Estimation}
We estimate the parameters of CMGND models using the ECM algorithm proposed by \cite{mgnd2025}, introducing new update equations for the constrained parameters as given in Eqs. (\ref{NRmu}), (\ref{NRsigma}) and (\ref{NRnu}). The updates of the unconstrained model can be found in \cite{mgnd2025,Wen2020,Bazi2006}.

From Eq. \ref{MGND} the log-likelihood function is given by 
\begin{equation}
\log L(\theta)=\sum_{n=1}^N\log{\Biggr[\sum_{k=1}^K\pi_kf_k(x_n|\mu_k, \sigma_k, \nu_k)\Biggr]}.
\label{loglikelihood}
\end{equation}

The \textbf{E-step} involves computing the following equation

\begin{equation}
Q(\theta,\theta^{(m-1)})=\sum_{n=1}^N\log{\Biggr[\sum_{k=1}^Kz_{nk}^{(m-1)}\pi_k^{(m-1)}f_k(x_n|\mu_k^{(m-1)}, \sigma_k^{(m-1)}, \nu_k^{(m-1)})\Biggr]},
\label{E-step}
\end{equation}

where 
\begin{equation*}
z_{nk}^{(m-1)}=\frac{\pi_k^{(m-1)}f_k(x_n|\mu_k^{(m-1)}, \sigma_k^{(m-1)}, \nu_k^{(m-1)})}{\sum_{k=1}^K\pi_k^{(m-1)}f_k(x_n|\mu_k^{(m-1)}, \sigma_k^{(m-1)}, \nu_k^{(m-1)})}.
\end{equation*}
The term $z_{nk}^{(m-1)}$ represents the current estimate of the posterior probability or responsibility \citep{Bishop2006} at the $(m-1)$-th iteration, that is, the probability that the observation $n$ belongs to the group $k$ given the parameters of the current component $\theta^{(m-1)}$.  

The parameter estimation in the $m$-th iteration is obtained by maximising $Q(\theta,\theta^{(m-1)})$ with respect to $\theta$, thus increasing the expectation of the complete likelihood of the data. In what follows,  we provide the updating equations for the parameters of the CMGND models.

\paragraph{Mixture weights} 
Set $\frac{\partial Q(\theta,\theta^{(m-1)})}{\partial\pi_k}=0$, then
\begin{equation}\label{Mweights}
\pi_k^{(m)}=\frac{\sum_{n=1}^Nz_{kn}^{(m-1)}}{\sum_{k=1}^K\sum_{n=1}^Nz_{kn}^{(m-1)}}.
\end{equation}

\paragraph{Location parameter}

To obtain the iteration equation for the constrained location parameter at any partition $r \in 1,2,...R$, where $\mu_k=\mu_r$ for all $k \in C_r$, we impose that the first derivative of the Q-function with respect to $\mu_r$ is set to zero, {\it i.e.}
\begin{equation*}
\begin{split}
\frac{\partial Q(\theta,\theta^{(m-1)})}{\partial \mu_r}&=\sum_{k \in C_r}\Biggr[\frac{\nu_k^{(m-1)}}{\bigr(\sigma_k^{(m-1)}\bigr)^{\nu^{(m-1)}_k}}\Biggr(\sum_{x_n\geq\mu_r^{(m-1)}}^Nz_{kn}^{(m-1)}(x_n-\mu_r^{(m-1)})^{\nu_k^{(m-1)}-1}\\
&-\sum_{x_n<\mu_r^{(m-1)}}^Nz_{kn}^{(m-1)}(\mu_r^{(m-1)}-x_n)^{\nu^{(m-1)}_k-1}\Biggr)\Biggr]=0.
\end{split}
\end{equation*}
Since the above equation is non-linear, the iterative Newton-Raphson method is applied as follows:
\begin{equation}\label{NRmu}
\mu_r^{(m)}=\mu_r^{(m-1)}-\frac{g\bigr(\mu_r^{(m-1)}\bigr)}{g'\bigr(\mu_r^{(m-1)}\bigr)},
\end{equation}
where
\begin{equation*}
\begin{split}
&g\bigr(\mu_r^{(m-1)}\bigr)=\sum_{k \in C_r}\Biggr[\frac{\nu^{(m-1)}_k}{\bigr(\sigma_k^{(m-1)}\bigr)^{\nu^{(m-1)}_k}}\Biggr(\sum_{x_n\geq \mu_r^{(m-1)}}^Nz_{kn}^{(m-1)}(x_n-\mu_r^{(m-1)})^{\nu^{(m-1)}_k-1}\\
&\hspace{2cm}-\sum_{x_n< \mu_r^{(m-1)}}^Nz_{kn}^{(m-1)}(\mu_r^{(m-1)}-x_n)^{\nu^{(m-1)}_k-1}\Biggr)\Biggr],\\
&g'\bigr(\mu_r^{(m-1)}\bigr)=\sum_{k \in C_r}\Biggr[-\frac{\nu^{(m-1)}_k}{\bigr(\sigma_k^{(m-1)}\bigr)^{\nu^{(m-1)}_k}}\Biggr(\sum_{x_n\geq\mu_r^{(m-1)}}^Nz_{kn}^{(m-1)}(x_n-\mu_r^{(m-1)})^{\nu^{(m-1)}_k-2}\\
&\hspace{2cm}\times(\nu^{(m-1)}_k-1)+\sum_{x_n<\mu_r^{(m-1)}}^Nz_{kn}^{(m-1)}(\mu_r^{(m-1)}-x_n)^{\nu^{(m-1)}_k-2}(\nu^{(m-1)}_k-1)\Biggr)\Biggr].\\
\end{split}
\end{equation*}
I removed the further derivation of eq. (6) see the commented code.

\paragraph{Scale parameter}
To obtain the iteration equation for the constrained scale parameter at any partition $r \in 1,2,...R$, where $\sigma_k=\sigma_r$ for all $k \in C_r$, we impose that the first derivative of the Q-function with respect to $\sigma_r$ is set to zero, {\it i.e.}
\begin{equation*}\label{sigmafpartial0}
\begin{split}
\frac{\partial Q(\theta,\theta^{(m-1)})}{\partial \sigma_r}&=\sum_{k \in C_r}\Biggr[\sum_{n=1}^Nz_{kn}^{(m-1)}\biggr(-\frac{1}{\sigma_r^{(m-1)}}\biggr)+\frac{\nu_k^{(m-1)}}{(\sigma_r^{(m-1)})^{\nu_k^{(m-1)}+1}}\sum_{n=1}^Nz_{kn}^{(m-1)}\bigr|x_n-\mu_k^{(m)}\bigr|^{\nu_k^{(m-1)}}\Biggr]=0.
\end{split}
\end{equation*}
The iterative Newton-Raphson method is applied as follows:
\begin{equation}\label{NRsigma}
\sigma_r^{(m)}=\sigma_r^{(m-1)}-\frac{g\bigr(\sigma_r^{(m-1)}\bigr)}{g'\bigr(\sigma_r^{(m-1)}\bigr)} ,
\end{equation}
where
\begin{equation*}
\begin{split}
&g\bigr(\sigma_r^{(m-1)}\bigr)=\sum_{k \in C_r}\Biggr[\sum_{n=1}^Nz_{kn}^{(m-1)}\biggr(-\frac{1}{\sigma_r^{(m-1)}}\biggr)+\frac{\nu^{(m-1)}_k}{\bigr(\sigma_r^{(m-1)}\bigr)^{\nu^{(m-1)}_k+1}}\sum_{n=1}^Nz_{kn}^{(m-1)}\bigr|x_n-\mu_k^{(m)}\bigr|^{\nu_k^{(m-1)}}\Biggr],\\
&g'\bigr(\sigma_r^{(m-1)}\bigr)=\sum_{k \in Cr}\Biggr[\sum_{n=1}^Nz_{kn}^{(m-1)}\frac{1}{\bigr(\sigma_r^{(m-1)}\bigr)^2}+\nu_k^{(m-1)}\bigr(-\nu_k^{(m-1)}-1\bigr)\\
&\hspace{2.5cm}\times\bigr(\sigma_r^{(m-1)}\bigr)^{-\nu_k^{(m-1)}-2}\sum_{n=1}^Nz_{kn}^{(m-1)}\bigr|x_n-\mu_k^{(m)}\bigr|^{\nu_k^{(m-1)}}\Biggr].
\end{split}
\end{equation*}

\paragraph{Shape parameter}
To obtain the iteration equation for the constrained shape parameter at any partition $r \in 1,2,...R$, where $\nu_k=\nu_r$ for all $k \in C_r$, we impose that the first derivative of the Q-function with respect to $\nu_r$ is set to zero, {\it i.e.}
\begin{equation*}
\begin{split}
\frac{\partial Q(\theta,\theta^{(m-1)})}{\partial \nu_r}&=\sum_{k \in C_r}\Biggr[\sum_{n=1}^Nz_{kn}^{(m-1)}\frac{1}{\nu_r^{(m-1)}}\Biggr(\frac{1}{\nu_r^{(m-1)}}\Psi\Biggr(\frac{1}{\nu_r^{(m-1)}}\Biggr)+1\Biggr)\\
&-\sum_{n=1}^Nz_{kn}^{(m-1)}\Biggr|\frac{x_n-\mu_k^{(m)}}{\sigma_k^{(m)}}\Biggr|^{\nu_r^{(m-1)}} \log \Biggr|\frac{x_n-\mu_k^{(m)}}{\sigma_k^{(m)}}\Biggr|\Biggr]=0.
\end{split}
\end{equation*}
Since the above equation is a non-linear equation, the iterative Newton-Raphson method is applied as follows:
\begin{equation}\label{NRnu}
\nu_r^{(m)}=\nu_r^{(m-1)}-\alpha\bigr(\nu_r^{(m-1)}\bigr)\frac{g\bigr(\nu_r^{(m-1)}\bigr)}{g'\bigr(\nu_r^{(m-1)}\bigr)},
\end{equation}
where: 
\begin{equation*}\label{nuf}
\begin{split}
&\alpha\bigr(\nu_r^{(m-1)}\bigr)=e^{-\nu_r^{(m-1)}},\\
&g\bigr(\nu_r^{(m-1)}\bigr)=\sum_{k \in C_r}\Biggr[\sum_{n=1}^Nz_{kn}^{(m-1)}\frac{1}{\nu_r^{(m-1)}}\Biggr(\frac{1}{\nu_r^{(m-1)}}\Psi\Biggr(\frac{1}{\nu_r^{(m-1)}}\Biggr)+1\Biggr)\\
&\hspace{2cm}-\sum_{n=1}^Nz_{kn}^{(m-1)}\Biggr|\frac{x_n-\mu^{(m)}_k}{\sigma^{(m)}_k}\Biggr|^{\nu_r^{(m-1)}} \log \Biggr|\frac{x_n-\mu^{(m)}_k}{\sigma^{(m)}_k}\Biggr|\Biggr],\\
&g'\bigr(\nu_r^{(m-1)}\bigr)=\sum_{k \in C_r}\Biggr[\sum_{n=1}^Nz_{kn}^{(m-1)}-\frac{1}{\bigr(\nu_r^{(m-1)}\bigr)^{2}}\Biggr(1+\frac{2}{\nu_r^{(m-1)}}\Psi\Biggr(\frac{1}{\nu_r^{(m-1)}}\Biggr)\\
&\hspace{2cm}+\frac{1}{\bigr(\nu_r^{(m-1)}\bigr)^{2}}\Psi'\Biggr(\frac{1}{\nu_r^{(m-1)}}\Biggr)\Biggr)-\sum_{n=1}^Nz_{kn}^{(m-1)}\Biggr|\frac{x_n-\mu^{(m)}_k}{\sigma^{(m)}_k}\Biggr|^{\nu_r^{(m-1)}}\Biggr(\log \Biggr|\frac{x_n-\mu^{(m)}_k}{\sigma^{(m)}_k}\Biggr|\Biggr)^2\Biggr].
\end{split}
\end{equation*}
The term $\alpha\bigr(\nu_r^{(m-1)}\bigr)=e^{-\nu_r^{(m-1)}}$ represents the adaptive step size function proposed by \cite{DuttiloR2024}, while a \textit{plain} updating equation can be recovered by setting $\alpha\bigr(\nu_r^{(m-1)}\bigr)=1$. The digamma $\Psi\bigr(1/\nu_r^{(m-1)}\bigr)$ and trigamma $\Psi'\bigr(1/\nu_r^{(m-1)}\bigr)$ functions are defined as follows
\begin{equation*}
\Psi\bigr(1/\nu_r^{(m-1)}\bigr)=\frac{\partial\Gamma\bigr(1/\nu_r^{(m-1)}\bigr)}{\partial\bigr(1/\nu_r^{(m-1)}\bigr)}\log\Gamma\bigr(1/\nu_r^{(m-1)}\bigr),\hspace{0.2cm} \Psi'\bigr(1/\nu_r^{(m-1)}\bigr)=\frac{\partial^2\Gamma\bigr(1/\nu_r^{(m-1)}\bigr)}{\partial\bigr(1/\nu_r^{(m-1)}\bigr)^2}\log\Gamma\bigr(1/\nu_r^{(m-1)}\bigr).
\end{equation*}

\section{Simulation study}\label{SM}
In the simulation, two important aspects are investigated: the performance of the parameter estimation in the presence of constraints and the choice among different constrained models. Two different sample sizes ($N=400, 1000$) and three levels of overlap between the mixture components (low, medium, and high) are considered.  
Data were simulated from the MGND and CMGND models with $K=3$ components and weights $\pi_1=0.4$, $\pi_2=0.3$ and $\pi_3=0.3$. We consider constraints that involve equality between two of the three components for the scale (UCU), the shape (UUC) and both the scale and shape (UCC) parameters.
Table \ref{simvalues} shows the parameters in the low overlap setting. The degree of overlap between the component densities is controlled by modifying the distances between the component means while keeping the other parameters constant. In particular, the component means are $\mu_1=0$, $\mu_2=7$ and $\mu_3=14$ for the medium overlap scenario and $\mu_1=0$, $\mu_2=5$ and $\mu_3=10$ for the high overlap scenario.    

\begin{table}[H]
\begin{center}
	\caption{MGND and CMGND simulation parameters.}\label{simvalues}
	\begin{tabular}{lcccc}
		\toprule
		\multicolumn{1}{l}{$\theta$}&\multicolumn{1}{c}{UUU}&\multicolumn{1}{c}{UCU}&\multicolumn{1}{c}{UUC}&\multicolumn{1}{c}{UCC}\\
		\midrule
		$\mu_1$  & 0  & 0 &0 & 0\\
		$\mu_2$ & 10  & 10 & 10 &  10\\
		$\mu_3$ & 20  & 20 & 20 &  20\\
		$\sigma_1$  & 0.2 & 0.2 & 0.2 & 0.2\\
		$\sigma_2$  & 1.5 & 3 &1.5 & 3\\
		$\sigma_3$  & 3 & 3 &3 & 3\\
		$\nu_1$  & 0.5 & 0.5 &0.5  &0.5\\
		$\nu_2$ & 1.6 & 1.6 &1.6 &  1.6\\
		$\nu_3$ & 4 & 4 &1.6 &  1.6\\
		\bottomrule
		\end{tabular}
\end{center}	
\end{table}

\begin{figure}[H]
\centering
	\includegraphics[width=14cm]{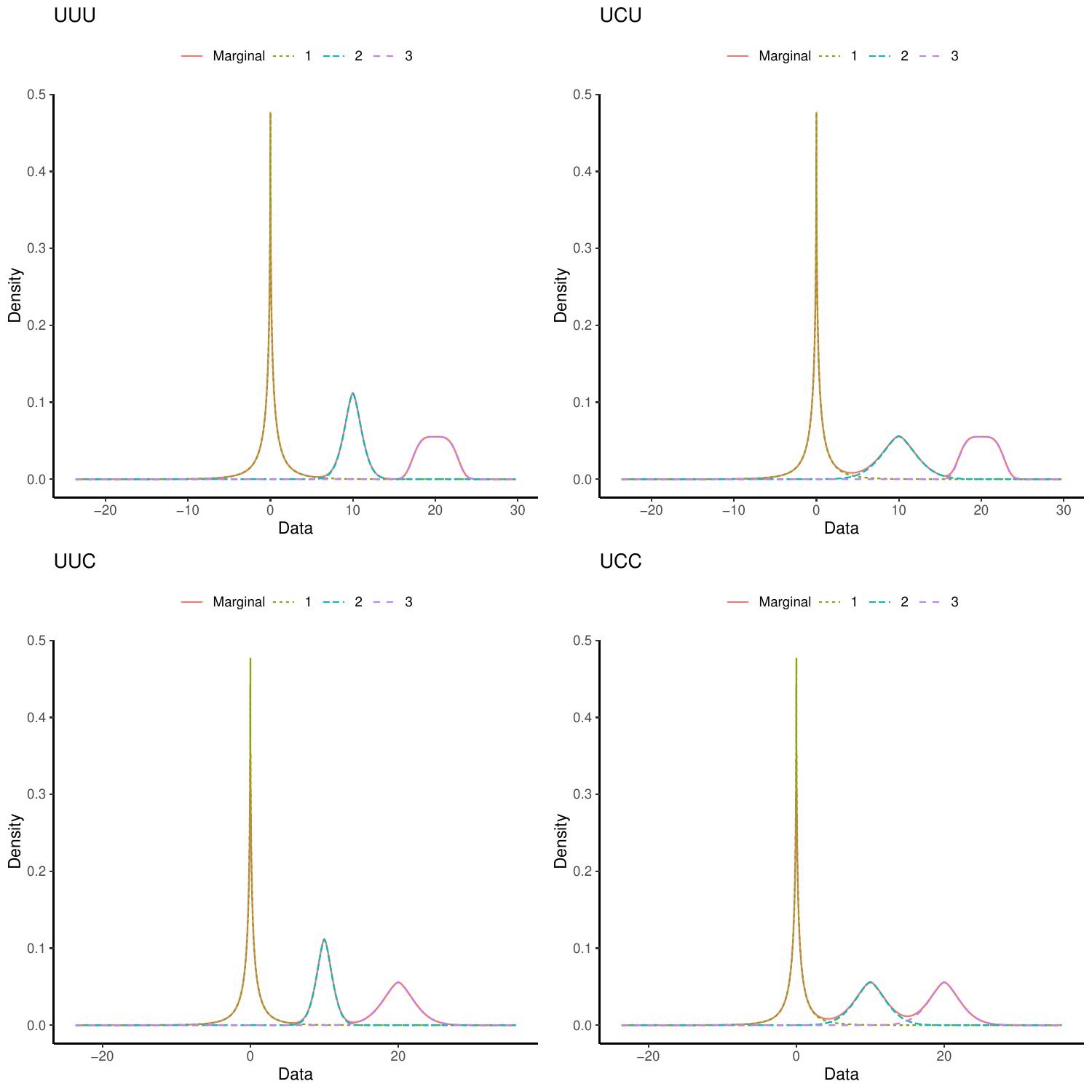}
	\caption{Simulated MGND and CMGND models: low overlap scenario.}
	\label{densityfiglo}
\end{figure}

\begin{figure}[H]
\centering
	\includegraphics[width=15cm]{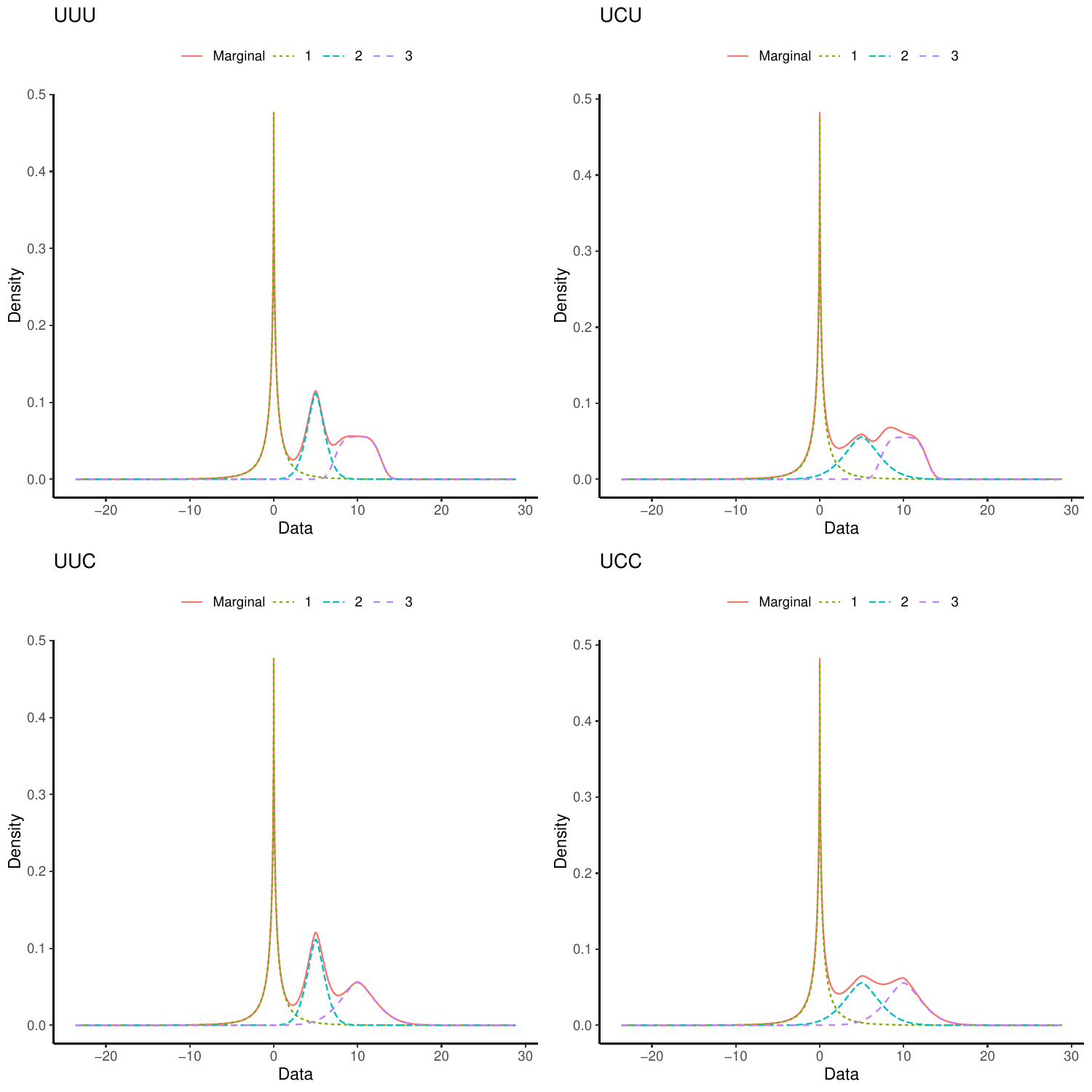}
	\caption{Simulated MGND and CMGND models: medium overlap scenario.}
	\label{densityfigmo}
\end{figure}

\begin{figure}[H]
\centering
	\includegraphics[width=15cm]{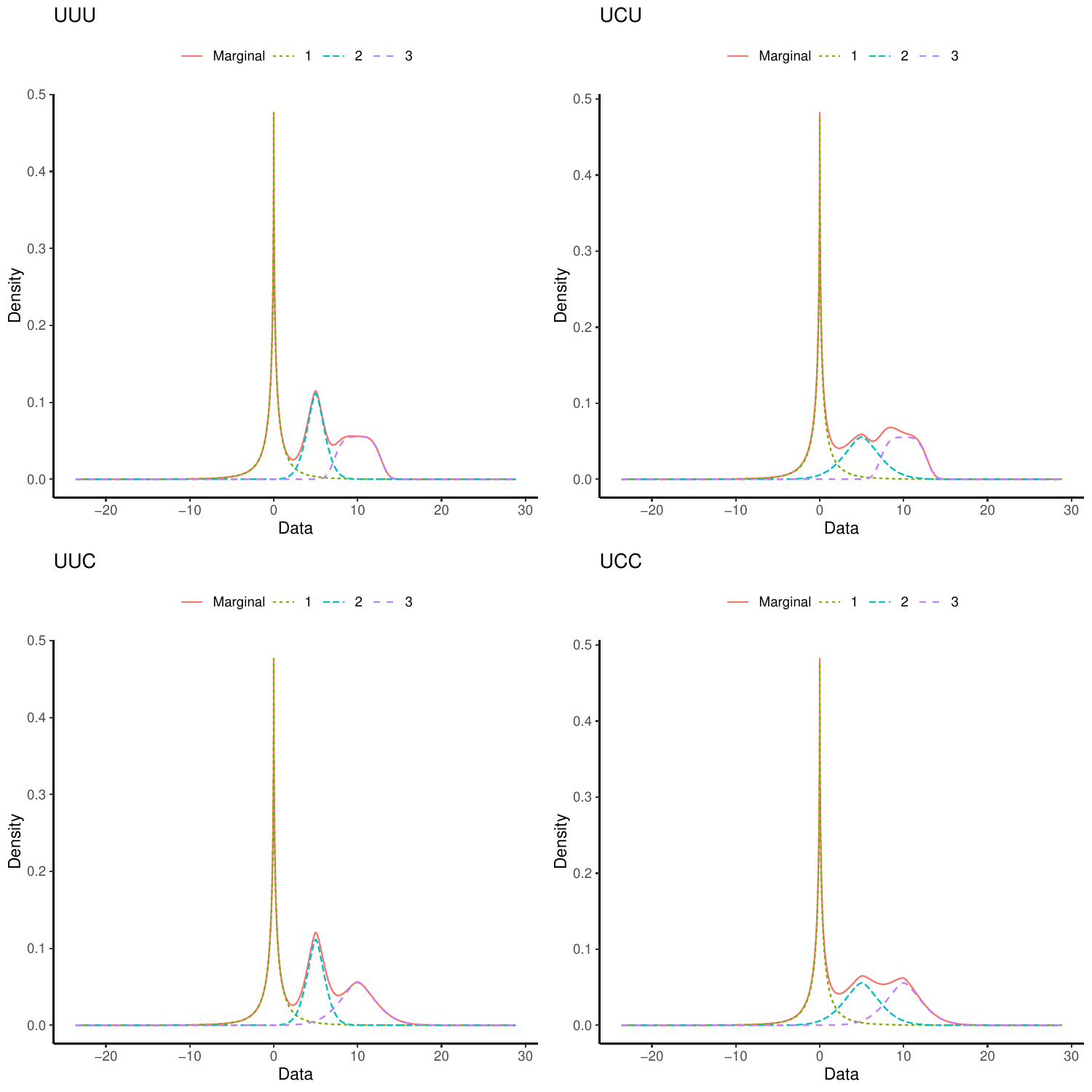}
	\caption{Simulated MGND and CMGND models: high overlap scenario.}
	\label{densityfigho}
\end{figure}

Figures \ref{densityfiglo}, \ref{densityfigmo}, and \ref{densityfigho} illustrate the density functions of the models used in the simulation study, highlighting the different constraints imposed on each model under different degree of overlap among the components of the mixture.

\subsection{Evaluation of parameter estimation in CMGND models}
We proceed by generating samples from three different CMGND models: 
\begin{itemize}
\item UCU: constraint on the scale for components $2$ and $3$ ($\sigma_2 = \sigma_3$), while the other parameters are unconstrained;
\item UUC: constraint on the shape for components $2$ and $3$ ($\nu_2 = \nu_3$), while the other parameters are unconstrained;
\item UCC: constraint on both scale and shape for components $2$ and $3$ ($\sigma_2 = \sigma_3$ and $\nu_2 = \nu_3$), while the other parameters are unconstrained.
\end{itemize}

The data is then fitted according to the constraints imposed during the data generation process. To assess estimation quality, the root mean square error (RMSE) is computed for each parameter, evaluating the ECMs algorithm's ability to correctly handle parameter constraints. The unconstrained MGND model is fitted to the same data to evaluate and compare the resulting RMSE values.

Results from the CMGND and MGND models in $250$ simulations  are reported for two different sample sizes ($N=400$ and $N=1000$) across three levels of overlap: low (Table \ref{RMSEl}), medium (Table \ref{RMSEm}) and high (Table \ref{RMSEh}).

When examining the RMSE values in the low overlap scenario, we note that for the first mixture component, whose parameters remain unconstrained, the RMSE values are quite similar between the unconstrained (MGND) and the constrained (CMGND) models. However, it is crucial to note the advantages in precision observed in the CMGND model for the parameters of the components involved in the constraints. For example, in the UCU model, the constrained parameter $\sigma_2$ has a significantly lower RMSE in the CMGND ($0.2$ with $N=400$ and $0.13$ with $N=1000$) compared to the MGND ($0.51$ with $N=400$ and $0.36$ with $N=1000$). Even more interestingly, an advantage in precision is also observed for the estimation of the shape parameter $\nu_2$, which is not constrained. This apparent paradox can be explained by considering the interdependence of parameter estimates within a mixture component, where in estimating $\nu_2$, the algorithm also uses information about the scale parameter $\sigma_2$. Since the CMGND-UCU model provides a more efficient estimate of $\sigma_2$, this better estimate positively impacts the estimate of $\nu_2$ within the same second component. 

By analysing Table \ref{RMSEm}, it is observed that RMSE values tend to be generally higher in the medium overlap scenario compared to the low overlap scenario for both models (MGND and CMGND). This is intuitive as it becomes more challenging to distinguish the mixture components and estimate their parameters accurately when the distributions overlap more substantially. In this scenario, we observe the same patterns observed in Table \ref{RMSEl} with one notable difference: in the UCC model, the advantage of precision of CMGND with respect to the estimation of unconstrained parameters within the same constrained components extends to $\mu_2$ and $\mu_3$.

Finally, in the high overlap scenario (Table \ref{RMSEh}), the benefits of the constraints on the precision of constrained parameters become more pronounced, especially for the smaller sample size ($N=400$). More significant reductions are also observed in the estimation of unconstrained parameters within components that have constrained parameters, such as $\mu_2$ and $\mu_3$. For the UCC model, particularly with $N=400$, a significant increase in precision is observed for the estimation of almost all parameters compared to the unconstrained MGND model. This includes the parameter $\nu_1$ of the first component, which is not directly involved in the constraints. This strengthens the idea that under high overlap conditions, the greater parsimony introduced by the constraints in the UCC model can significantly stabilise the overall estimation process, leading to more accurate estimates even for unconstrained components. 

In conclusion, fitting a CMGND model that correctly reflects the underlying constraints generally leads to more accurate parameter estimates (lower RMSE) compared to fitting an unconstrained MGND model. This is particularly noticeable for the parameters that are actually constrained in the data-generation process. Although increasing sample size tends to reduce the gap between constrained and unconstrained models, CMGND models maintain competitive performance, emphasising their potential for more reliable inference in complex situations. 
These results support the idea that the incorporation of correct constraints in the model can enhance parameter estimation. As a consequence, a key aspect to consider is the ability of model selection criteria such as the Bayesian Information Criterion (BIC) to identify the correctly constrained model. This point will be investigated in the next subsection.

\begin{table}[H]
\caption{RMSE of parameter estimates for simulated constrained MGND models.}
\label{RMSEl}
\centering
\begin{tabular}{lrrrrrrrrrrrr}
\toprule
\multicolumn{12}{c}{Low overlap}\\
\midrule
&\multicolumn{1}{c}{$\pi_1$}&\multicolumn{1}{c}{$\mu_1$}&
\multicolumn{1}{c}{$\sigma_1$}&\multicolumn{1}{c}{$\nu_1$}&\multicolumn{1}{c}{$\pi_2$}&\multicolumn{1}{c}{$\mu_2$}&\multicolumn{1}{c}{$\sigma_2$}&\multicolumn{1}{c}{$\nu_2$}&\multicolumn{1}{c}{$\mu_3$}&\multicolumn{1}{c}{$\sigma_3$}&\multicolumn{1}{c}{$\nu_3$}&N\\
\midrule
\multicolumn{12}{c}{UCU}&400\\
$\text{CMGND}$ & 0.01 & 0.05 & 0.11 & 0.07 & 0.01 & 0.25 & 0.20 & 0.37& 0.15& 0.20&0.82 \\ 
$\text{MGND}$ & 0.01 & 0.05 & 0.11 & 0.07 & 0.01 & 0.25 & 0.51 & 0.58& 0.15& 0.21&0.84 \\ 
\multicolumn{12}{c}{}&1000\\
$\text{CMGND}$ & 0.01 & 0.03 & 0.08 & 0.05 & 0.01 & 0.16 & 0.13 & 0.21& 0.11 & 0.13& 0.64 \\ 
$\text{MGND}$ & 0.01 & 0.03 & 0.08 & 0.05 & 0.01 & 0.17 & 0.36 & 0.36& 0.11& 0.13 &0.65 \\ 
\midrule
\multicolumn{12}{c}{UUC}&400\\
$\text{CMGND}$ & 0.00 & 0.05 & 0.12 & 0.08 & 0.01 & 0.12 & 0.24 & 0.33& 0.23 & 0.46& 0.35 \\ 
$\text{MGND}$ & 0.00 & 0.05 & 0.11 & 0.07 & 0.01 & 0.12 & 0.25 & 0.46& 0.22& 0.49&0.49 \\ 
\multicolumn{12}{c}{}&1000\\
$\text{CMGND}$ & 0.00 & 0.03 & 0.08 & 0.05 & 0.00 & 0.08 & 0.15 & 0.19& 0.14 & 0.26& 0.18 \\ 
$\text{MGND}$ & 0.00 & 0.03 & 0.08 & 0.05 & 0.00 & 0.08 & 0.17 & 0.29& 0.15& 0.30&0.25 \\ 
\midrule
\multicolumn{12}{c}{UCC}&400\\
$\text{CMGND}$ & 0.01 & 0.05 & 0.11 & 0.07 & 0.01 & 0.25 & 0.36 & 0.36& 0.23 & 0.36& 0.36 \\ 
$\text{MGND}$ & 0.01 & 0.05 & 0.11 & 0.08 & 0.03 & 0.28 & 0.58 & 0.73& 0.28&0.52&0.56 \\ 
\multicolumn{12}{c}{}&1000\\
$\text{CMGND}$ & 0.00 & 0.03 & 0.09 & 0.05 & 0.01 & 0.17 & 0.23 & 0.20& 0.15 & 0.23& 0.20 \\ 
$\text{MGND}$ & 0.01 & 0.03 & 0.08 & 0.05 & 0.01 & 0.18 & 0.38 & 0.38& 0.17& 0.31&0.26 \\ 
\bottomrule
\end{tabular}
\end{table}

\begin{table}[H]
\caption{RMSE of parameter estimates for simulated constrained MGND models.}
\label{RMSEm}
\centering
\begin{tabular}{lrrrrrrrrrrrr}
\toprule
\multicolumn{12}{c}{Medium overlap}\\
\midrule
&\multicolumn{1}{c}{$\pi_1$}&\multicolumn{1}{c}{$\mu_1$}&
\multicolumn{1}{c}{$\sigma_1$}&\multicolumn{1}{c}{$\nu_1$}&\multicolumn{1}{c}{$\pi_2$}&\multicolumn{1}{c}{$\mu_2$}&\multicolumn{1}{c}{$\sigma_2$}&\multicolumn{1}{c}{$\nu_2$}&\multicolumn{1}{c}{$\mu_3$}&\multicolumn{1}{c}{$\sigma_3$}&\multicolumn{1}{c}{$\nu_3$}&N\\
\midrule
\multicolumn{12}{c}{UCU}&400\\
$\text{CMGND}$ & 0.02 & 0.05 & 0.12 & 0.08 & 0.03 & 0.32 & 0.39 & 0.70& 0.22& 0.22&0.94 \\ 
$\text{MGND}$ & 0.02 & 0.05 & 0.12 & 0.08 & 0.03 & 0.37 & 0.66 & 0.81& 0.25& 0.27&0.94 \\ 
\multicolumn{12}{c}{}&1000\\
$\text{CMGND}$ & 0.01 & 0.03 & 0.09 & 0.06 & 0.02 & 0.22 & 0.14 & 0.55& 0.15 & 0.14& 0.71 \\ 
$\text{MGND}$ & 0.01 & 0.03 & 0.09 & 0.06 & 0.03 & 0.22 & 0.40 & 0.54& 0.17& 0.16 &0.70 \\ 
\midrule
\multicolumn{12}{c}{UUC}&400\\
$\text{CMGND}$ & 0.01 & 0.05 & 0.12 & 0.08 & 0.03 & 0.12 & 0.33 & 0.46& 0.28 & 0.46& 0.40 \\ 
$\text{MGND}$ & 0.01 & 0.05 & 0.11 & 0.07 & 0.02 & 0.13 & 0.29 & 0.58& 0.27& 0.49&0.51 \\ 
\multicolumn{12}{c}{}&1000\\
$\text{CMGND}$ & 0.01 & 0.03 & 0.09 & 0.06 & 0.01 & 0.09 & 0.15 & 0.21& 0.17 & 0.29& 0.20 \\ 
$\text{MGND}$ & 0.01 & 0.03 & 0.08 & 0.05 & 0.01 & 0.09 & 0.19 & 0.35& 0.17& 0.31&0.26 \\ 
\midrule
\multicolumn{12}{c}{UCC}&400\\
$\text{CMGND}$ & 0.02 & 0.05 & 0.13 & 0.08 & 0.03 & 0.42 & 0.68 & 0.45& 0.31 & 0.53& 0.46 \\ 
$\text{MGND}$ & 0.04 & 0.05 & 0.14 & 0.21 & 0.07 & 0.69 & 1.01 & 0.84& 0.61&0.84&0.82 \\ 
\multicolumn{12}{c}{}&1000\\
$\text{CMGND}$ & 0.01 & 0.03 & 0.09 & 0.05 & 0.01 & 0.22 & 0.23 & 0.22& 0.18 & 0.23& 0.22 \\ 
$\text{MGND}$ & 0.02 & 0.03 & 0.09 & 0.07 & 0.05 & 0.28 & 0.56 & 0.59& 0.29& 0.39&0.32 \\ 
\bottomrule
\end{tabular}
\end{table}

\begin{table}[H]
\caption{RMSE of parameter estimates for simulated constrained MGND models.}
\label{RMSEh}
\centering
\begin{tabular}{lrrrrrrrrrrrr}
\toprule
\multicolumn{12}{c}{Hig overlap}\\
\midrule
&\multicolumn{1}{c}{$\pi_1$}&\multicolumn{1}{c}{$\mu_1$}&
\multicolumn{1}{c}{$\sigma_1$}&\multicolumn{1}{c}{$\nu_1$}&\multicolumn{1}{c}{$\pi_2$}&\multicolumn{1}{c}{$\mu_2$}&\multicolumn{1}{c}{$\sigma_2$}&\multicolumn{1}{c}{$\nu_2$}&\multicolumn{1}{c}{$\mu_3$}&\multicolumn{1}{c}{$\sigma_3$}&\multicolumn{1}{c}{$\nu_3$}&N\\
\midrule
\multicolumn{12}{c}{UCU}&400\\
$\text{CMGND}$ & 0.05 & 0.05 & 0.16 & 0.23 & 0.07 & 1.02 & 0.95 & 0.84& 0.57&0.71&1.21 \\ 
$\text{MGND}$ & 0.08 & 0.05 & 0.17 & 0.39 & 0.10 & 1.52 & 1.31 & 1.15& 0.80&0.76&1.15 \\ 
\multicolumn{12}{c}{}&1000\\
$\text{CMGND}$ & 0.03 & 0.03 & 0.10 & 0.07 & 0.03 & 0.42 & 0.34 & 0.77& 0.23 & 0.22& 0.73 \\ 
$\text{MGND}$ & 0.04 & 0.03 & 0.11 & 0.14 & 0.08 & 0.95 & 0.91 & 1.06& 0.58&0.53&0.82 \\ 
\midrule
\multicolumn{12}{c}{UUC}&400\\
$\text{CMGND}$ & 0.03 & 0.05 & 0.13 & 0.10 & 0.07 & 0.17 & 0.38 & 0.60& 0.57 & 0.76& 0.62 \\ 
$\text{MGND}$ & 0.03 & 0.05 & 0.12 & 0.10 & 0.07 & 0.18 & 0.36 & 0.81& 0.61&0.83&0.71 \\ 
\multicolumn{12}{c}{}&1000\\
$\text{CMGND}$ & 0.02 & 0.03 & 0.09 & 0.06 & 0.04 & 0.11 & 0.21 & 0.33& 0.35 & 0.47& 0.30 \\ 
$\text{MGND}$ & 0.01 & 0.03 & 0.09 & 0.06 & 0.04 & 0.11 & 0.22 & 0.52& 0.35&0.46&0.32 \\ 
\midrule
\multicolumn{12}{c}{UCC}&400\\
$\text{CMGND}$ & 0.04 & 0.05 & 0.15 & 0.18 & 0.08 & 1.01 & 1.09 & 0.66& 0.82 & 0.92& 0.62 \\ 
$\text{MGND}$ & 0.08 & 0.05 & 0.18 & 0.31 & 0.12 & 1.34 & 1.46 & 1.02& 1.00&1.22&0.92 \\ 
\multicolumn{12}{c}{}&1000\\
$\text{CMGND}$ & 0.02 & 0.03 & 0.10 & 0.06 & 0.05 & 0.41 & 0.62 & 0.29& 0.40 & 0.55& 0.28 \\ 
$\text{MGND}$ & 0.05 & 0.03 & 0.11 & 0.13 & 0.12 & 0.69 & 1.08 & 0.89& 0.71&0.76&0.65 \\ 
\bottomrule
\end{tabular}
\end{table}

\subsection{Model selection among constrained and unconstrained MGND models}
We proceed by generating data under different constraints and fitting all candidate models, including the unconstrained MGND model, to each generated dataset. For each model fit to the simulated dataset, the BIC is calculated as follows
\begin{equation}
\text{BIC}=p\log(N)-2\log L(\widehat{\theta}).
\label{eq.bic}
\end{equation}

\begin{figure}[H]
\centering
	\includegraphics[width=16cm]{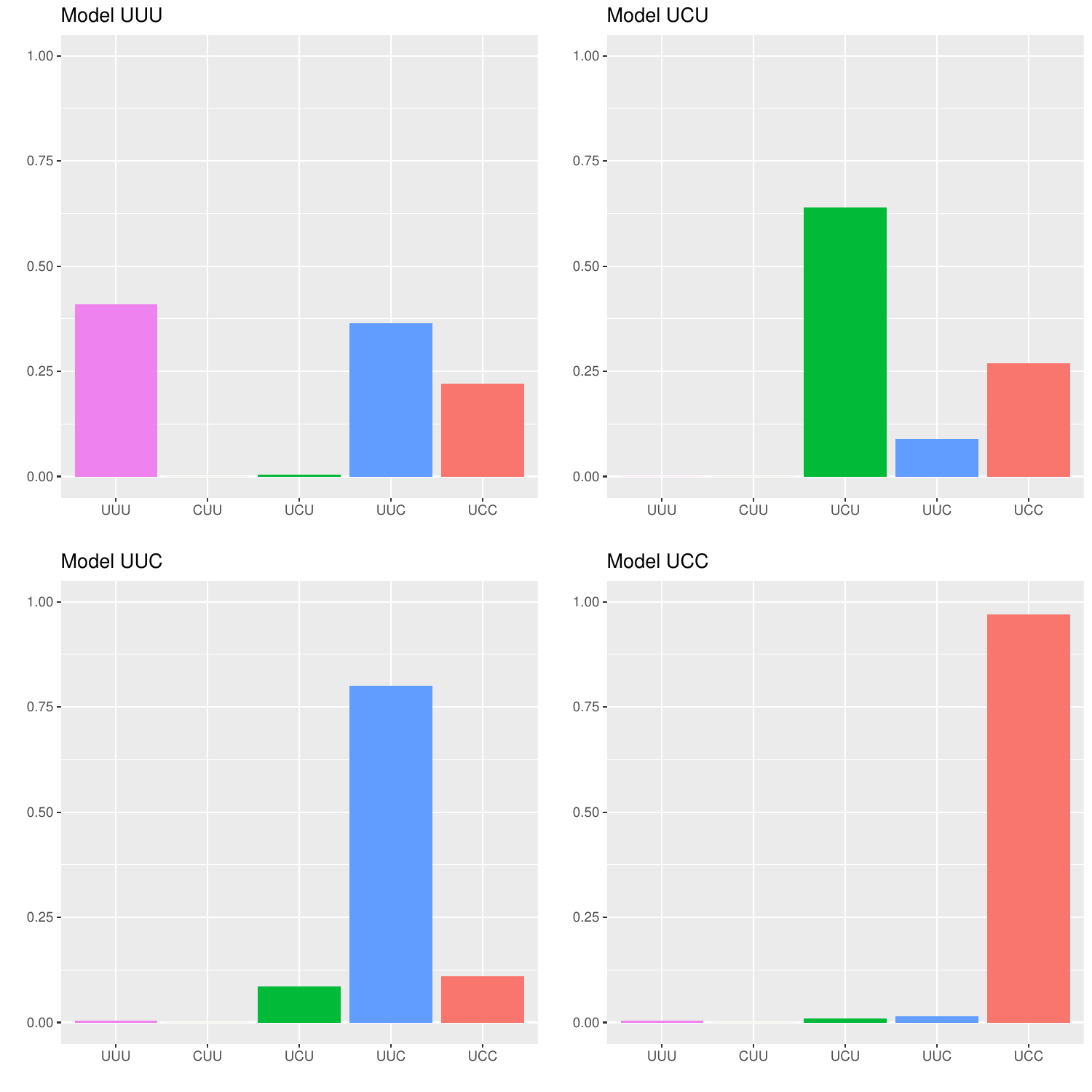}
	\caption{Proportion of times models selected by BIC: low overlap scenario. Sample size $N=400$.}
	\label{bic400lowo}
\end{figure}

\begin{figure}[H]
\centering
	\includegraphics[width=16cm]{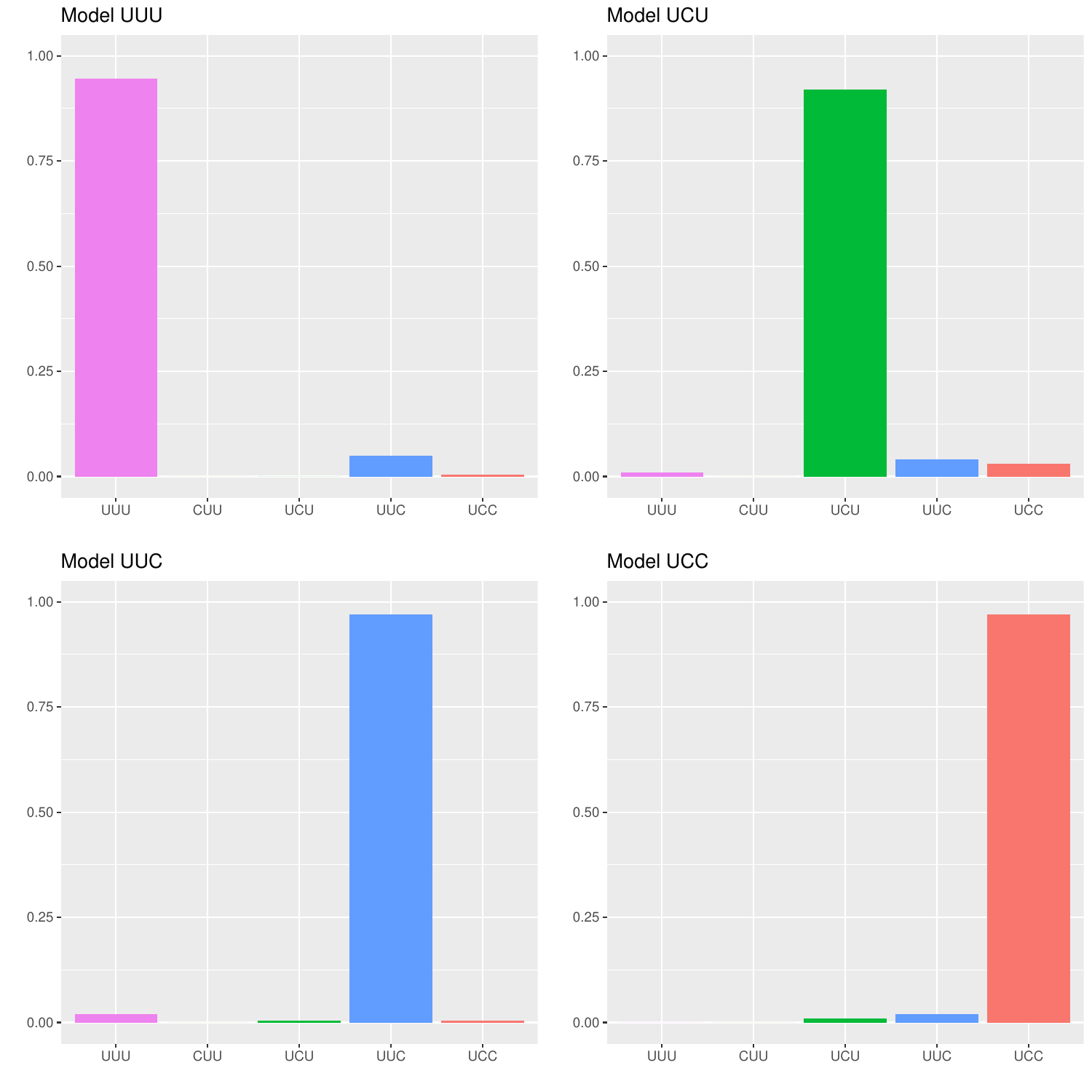}
	\caption{Proportion of times models selected by BIC: low overlap scenario. Sample size $N=1000$.}
	\label{bic1000lowo}
\end{figure}

\begin{figure}[H]
\centering
	\includegraphics[width=16cm]{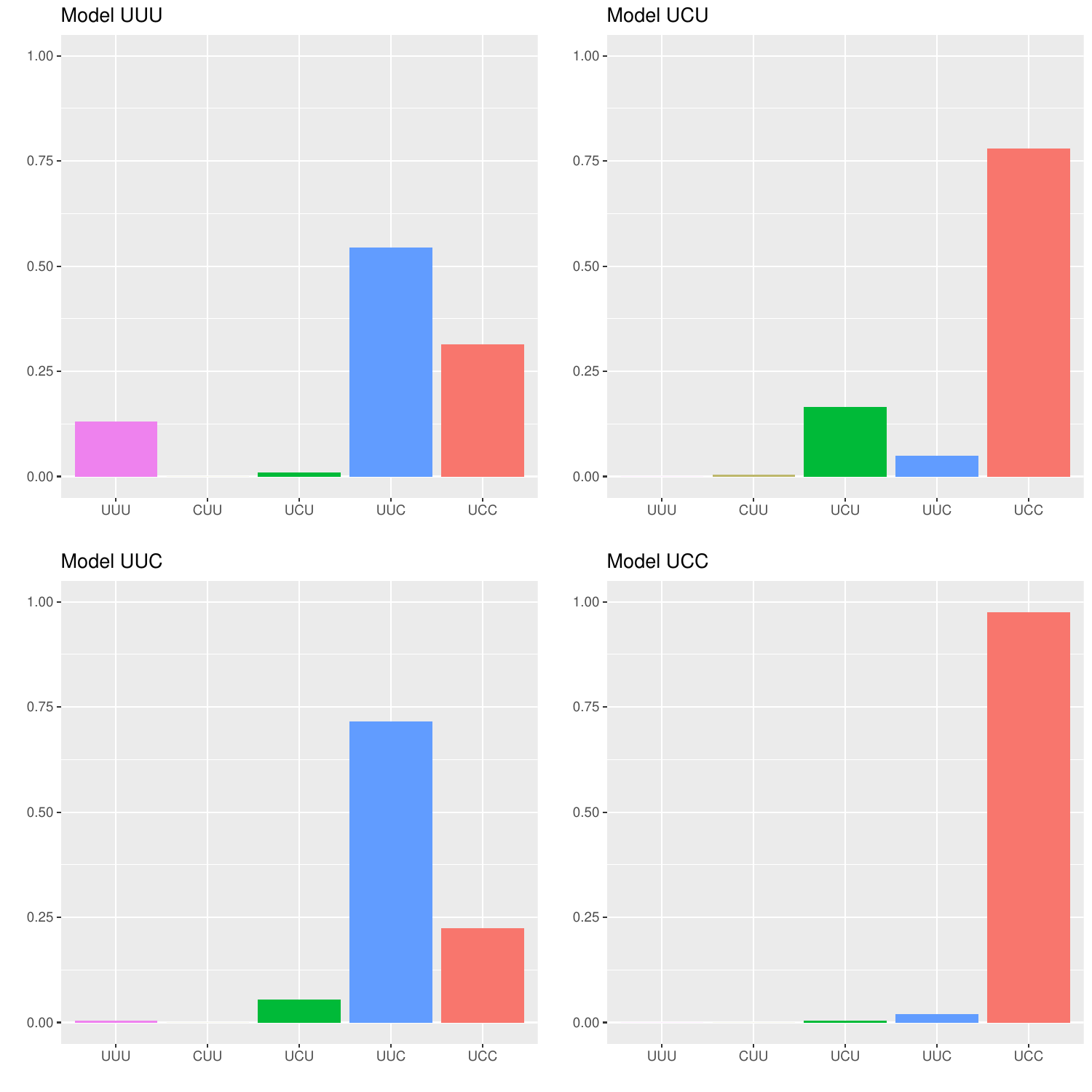}
	\caption{Proportion of times models selected by BIC: medium overlap scenario. Sample size $N=400$.}
	\label{bic400mediumo}
\end{figure}

\begin{figure}[H]
\centering
	\includegraphics[width=16cm]{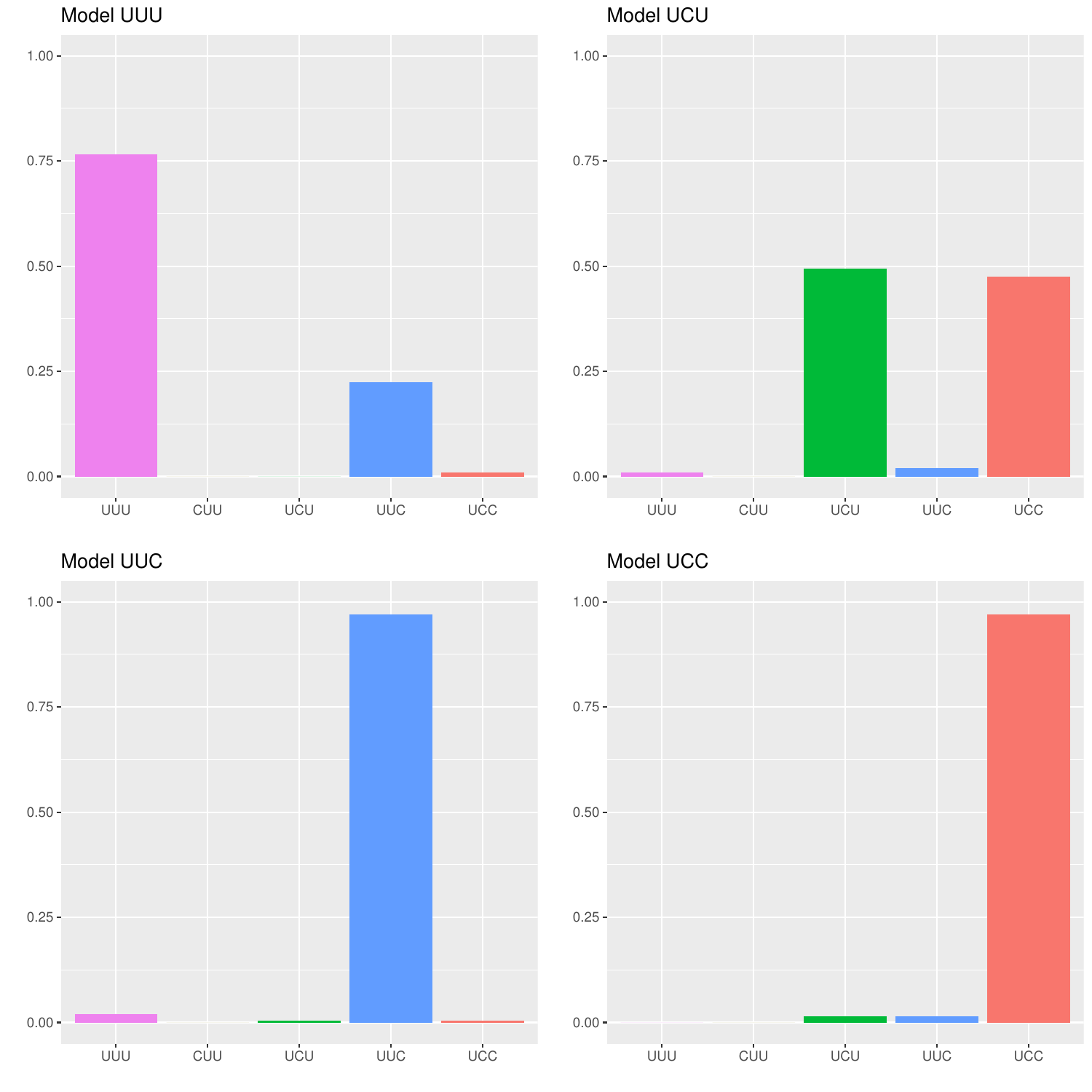}
	\caption{Proportion of times models selected by BIC: medium overlap scenario. Sample size $N=1000$.}
	\label{bic1000mediumo}
\end{figure}

\begin{figure}[H]
\centering
	\includegraphics[width=16cm]{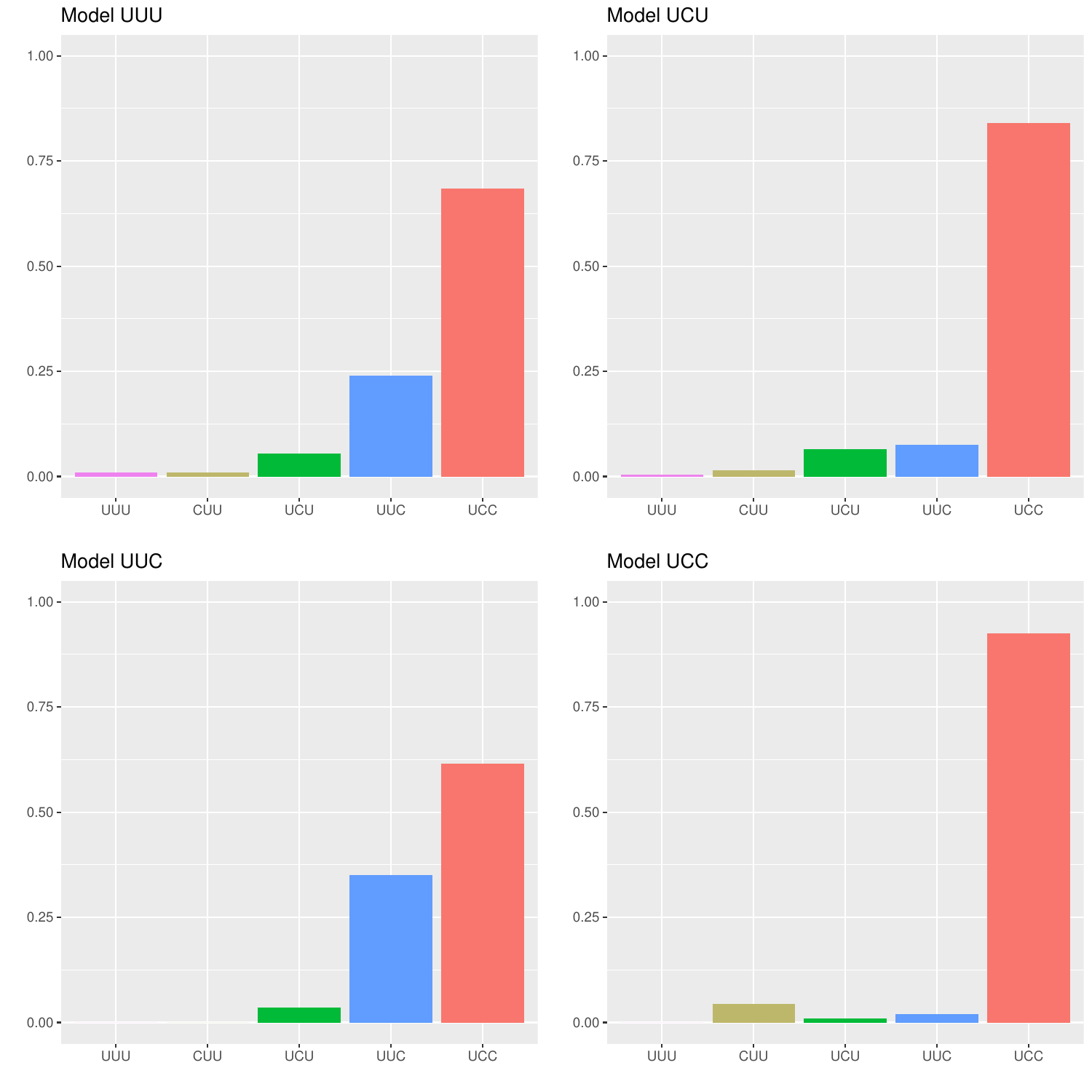}
	\caption{Proportion of times models selected by BIC: high overlap scenario. Sample size $N=400$.}
	\label{bic400higho}
\end{figure}

\begin{figure}[H]
\centering
	\includegraphics[width=16cm]{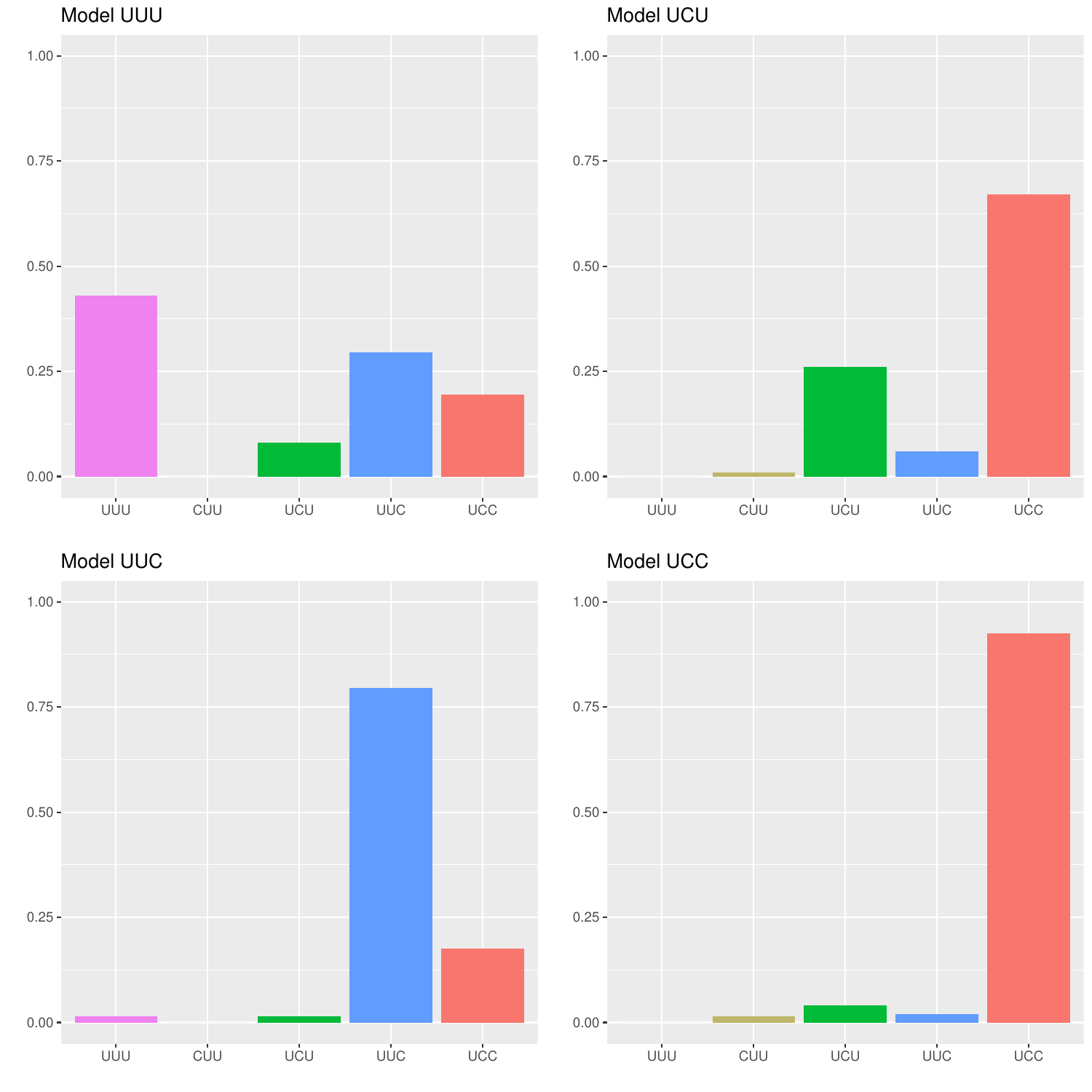}
	\caption{Proportion of times models selected by BIC: high overlap scenario. Sample size $N=1000$.}
	\label{bic1000higho}
\end{figure}

We examine how often the BIC criterion successfully identifies the constrained model responsible for generating the data. The results are shown in figures \ref{bic400lowo}-\ref{bic1000higho}.

In the low overlap scenario and small sample size ( figure \ref{bic400lowo}) BIC shows a good identification of the scale constraint ($64\%$) and the shape constraint ($80\%$). When data are generated from the UCC model, BIC predominantly selects the UCC model ($97\%$) demonstrating good recognition of both constraints. Instead, when data are generated from the unconstrained UUU model, BIC selects the UUU model only $41\%$ of times with a lot of instances of selecting constrained models, indicating some uncertainty when the sample size is limited. With a larger sample size, BIC's ability to select the correct model with and without constraints improves significantly.

In the medium overlap scenario (Figures \ref{bic400mediumo} and \ref{bic1000mediumo}), BIC's performance in selecting the true model slightly worsens compared to the low overlap scenario. 
In fact, especially with a sample size of $N=400$, the information in the data may not be sufficient to definitively discriminate between the true nature of the constraints. Notable is what happens when the true model is UUU and UCU. When the true model is UUU, BIC selects the UUC model $54.5\%$ of times and the UCC model $31.5\%$ of times, while the UUU model is chosen only $13\%$ of times. In a scenario of intermediate overlap and small sample size, it seems that the likelihood improvement achieved by the more complex model (UUU) is insufficient to compensate for the complexity penalty. At the same time, even though the true model is UUU, with a sample size of $N=400$, there may not be sufficient information in the data to accurately estimate all parameters of the unconstrained model. Furthermore, when the true model is UCU the BIC selects the UCC model $78\%$ a number of times, {\it i.e.} a model with an "incorrect" constraint structure might have a greater capacity to adapt to the observed characteristics of the sample. What has been observed is, in fact, a precursor to a more widespread phenomenon that occurs in the scenario with high overlap.

Under high overlap conditions, the difficulty in correctly selecting the true model using BIC becomes even more evident (figures \ref{bic400higho} and \ref{bic1000higho}). There is a clear tendency of BIC to select the more parsimonious UCC model, especially with $N=400$.
This is an indication that the UCC model offers a good compromise between goodness-of-fit and simplicity, particularly under challenging conditions for parameter identification. In high overlap situations and with a limited sample size like $N=400$, the complexity of a model with many parameters (such as the unconstrained MGND or models with fewer constraints like UCU or UUC) might not be justified by the improvement in log-likelihood it provides. 
There could also be identifiability issues when the distributions of the different mixture components are very close to each other, making it difficult for the estimation algorithm to determine which component each observation belongs to. Imposing constraints, as in UCC, reduces the parameter space and can help stabilise the estimation process, preventing spurious solutions, and improving identifiability.

Finally, the effect of sample size must be considered: with more data ($N=1000$), the BIC is generally more likely to correctly identify the true model in all overlap scenarios, although the extent of this improvement varies. In the low overlap scenario, the BIC's ability to correctly identify the true model improves significantly for all models (UUU, UCU, UUC, and UCC). In the intermediate overlap scenario, a notable improvement in the correct model selection is observed for UUU, UUC, and UCC models. In the high-overlapping scenario, only the UUC and UCC models are correctly identified by the BIC.

\subsection{Evaluation of BIC-selected models via RMSE of marginal distribution moments}
Following the selection of candidate models based on the BIC, this section aims to provide a further evaluation of the chosen models by assessing their ability to capture the essential characteristics of the marginal data distribution. 
The analysis will focus on the first four central moments: the mean, the variance, the skewness, and the kurtosis. These moments provide a comprehensive summary of the distribution's shape beyond just its location and scale. 

For each model under consideration, we derive the theoretical expressions for the mean, variance, skewness, and kurtosis of the marginal distribution. For mixture of generalised normal distribution models, the formulas for these moments depend on the parameters of each component (weight $\pi_k$, location $\mu_k$, scale $\sigma_k$, and shape $\nu_k$). We refer to \cite{mgnd2025} for the analytical expression of the first four moments of the marginal distribution of the MGND.
Then, for each simulated data, we use the mixture estimated parameters to compute an estimate of the theoretical moments implied by the model and compare them to the original ones. Finally, we will compute the RMSE for each of these four moments, comparing the theoretical values predicted by the model and the moments estimated from the data by the BIC-selected model. 
A lower RMSE for a particular moment will indicate a better ability of the model to reproduce that specific characteristic of the data's marginal distribution.
For comparison, we will compute the RMSE of the estimates obtained by the model corresponding to the true generating process.
By examining the RMSE of these marginal moments, we aim to gain a deeper understanding of the adequacy of the BIC-selected models in capturing the fundamental properties of the data distribution, complementing the likelihood-based model selection process. 
Table \ref{case1} compares the RMSE for the first four central moments between the true model that generates the data and the models most frequently selected by BIC under various overlap conditions and sample sizes. In particular, the following scenarios are investigated:
\begin{itemize}
\item low and intermediate overlap, $N=400$, true model: UUU. BIC's selected models: UUC and UCC;
\item intermediate overlap, $N=400$ and $1000$, true model: UCU. BIC's selected model: UCC;
\item high overlap, $N=400$, true models: UUU, UCU, UUC. BIC's selected model: UCC;
\item high overlap, $N=1000$, true models: UUU. BIC's selected models: UUC and UCC;
\item high overlap, $N=1000$, true models: UCU. BIC's selected model: UCC;
\end{itemize}

With low overlap and small sample size, when the true model is UUU constrained models UUC and UCC provide estimates for the marginal moments that are not far from the true model's estimates, although there is a slight loss of accuracy in variance estimation. 
In the case of intermediate overlap, both for the true model UUU and the true model UCU, the UCC model shows comparable or even better capability (for kurtosis when $N=400$) to capture the marginal moments. With high overlap, the same pattern is observed even with a large sample size ($N=1000$).
In summary, the BIC's selected models, even when not matching the true data-generating process, tend to provide reasonable estimates for the first four moments of the marginal distribution.
In particular, the UCC model can exhibit comparable or even better adaptability to identify key characteristics of the distribution, particularly kurtosis. This suggests that in complex scenarios, simpler models offer a better trade-off between goodness-of-fit and parsimony, avoiding potential overfitting or instability in estimating more complex models.
In summary, the analysis of Table 6 supports the idea that in scenarios of high overlap and/or limited sample sizes, simpler models can still deliver valid and sometimes superior results.

\begin{table}[H]
\caption{RMSE of marginal distribution parameter estimates.}
\label{case1}
\centering
\begin{tabular}{lccccc}
\toprule
Overlap & $\mu$ & $\sigma^2$ & skewness & kurtosis & N \\
\hline
Low & \multicolumn{4}{c}{UUU} & 400\\
\hline
UUU & 0.071 & 1.169 & 0.033 & 0.096 \\
UUC & 0.086 & 1.350 & 0.034 & 0.109 \\
UCC & 0.078 & 1.314 & 0.034 & 0.100 \\
\hline
Intermediate & \multicolumn{4}{c}{UUU} & 400\\
\hline
UUU & 0.079 & 1.006 & 0.060 & 0.293 \\
UUC & 0.091 & 1.096 & 0.065 & 0.312 \\
UCC & 0.079 & 1.044 & 0.059 & 0.240 \\
\toprule
Intermediate & \multicolumn{4}{c}{UCU} & 400\\
\hline
UCU & 0.102 & 1.161 & 0.065 & 0.311 \\
UCC & 0.109 & 1.204 & 0.067 & 0.282 \\
\hline
Intermediate & \multicolumn{4}{c}{UCU} & 1000\\
\hline
UCU & 0.066 & 0.822 & 0.042 & 0.196 \\
UCC & 0.071 & 0.881 & 0.042 & 0.193 \\
\hline
High & \multicolumn{4}{c}{UUU} & 400\\
\hline
UUU & 0.088 & 0.904 & 0.104 & 0.971 \\
UCC & 0.083 & 0.858 & 0.093 & 0.713 \\
\hline
High & \multicolumn{4}{c}{UCU} & 400\\
\hline
UCU & 0.105 & 0.966 & 0.102 & 0.953 \\
UCC & 0.109 & 0.995 & 0.098 & 0.590 \\
\hline
High & \multicolumn{4}{c}{UUC} & 400\\
\hline
UUC & 0.106 & 1.072 & 0.124 & 0.866 \\
UCC & 0.106 & 1.055 & 0.138 & 0.849 \\
\hline
High & \multicolumn{4}{c}{UUU} & 1000\\
\hline
UUU & 0.051 & 0.635 & 0.067 & 0.556 \\
UUC & 0.067 & 0.653 & 0.086 & 0.519 \\
UCC & 0.052 & 0.643 & 0.071 & 0.530 \\
\hline
High & \multicolumn{4}{c}{UCU} & 1000\\
\hline
UCU & 0.065 & 0.647 & 0.066 & 0.436 \\
UCC & 0.064 & 0.624 & 0.059 & 0.368 \\
\bottomrule
\end{tabular}
\end{table}

\section{Real data application}\label{RDA}
This section presents a comparative analysis on the 50 constituents of the Euro Stoxx 50 index (SX5E). The goodness-of-fit of the two-component CMGND model is compared to that of the two-component constrained mixture of normals (CMND) and the two-component constrained mixture of Student-t distributions (CMSTD). The two-component CMGND model is estimated using the proposed algorithm ECMs, while CMND and CMSTD are estimated with the packages R \textit{mclust} \citep{Scrucca2023}, and \textit{teigen} \citep{Andrews2011,Andrews2018}. While \textit{mclust} enables the estimation of constrained mixtures of normal distributions with equal variance, \textit{teigen} allows for the estimation of constrained mixtures of Student-\textit{t} distributions with common scale and/or shape parameters. All algorithms employ 5 starting points with k-means initialisation.

Data on the daily closing prices of the 50 constituents of the SX5E have been collected from Refinitiv \citep{LSEGData2025} for the period from January 4, 2010, to September 30, 2024. Daily log-returns for each stock are computed using the natural log-difference approach as follows
\begin{equation}
    r_{t}=\ln(P_t-P_{t-1})100,
\end{equation}
where: $r_{t}$ is the daily percentage return of the equity index at time $t$, $P_{t}$ is the daily closing price of the equity index at time $t$ and $P_{t-1}$ is the daily closing price of the equity index at time $t-1$.

The descriptive statistics and the JB test are reported in Table \ref{tab.comp.bs} in Appendix \ref{sec.appendix}. All stocks exhibit a mean and median close to zero. The standard deviation ranges from 1.2370 to 3.1286. Returns are predominantly negatively skewed, with an empirical kurtosis exceeding three, indicating fat-tailed distributions. The JB test rejects the null hypothesis of normality for all stocks. The models' performance is assessed using the BIC.

\begin{figure}[H]
\centering
	\includegraphics[width=12cm]{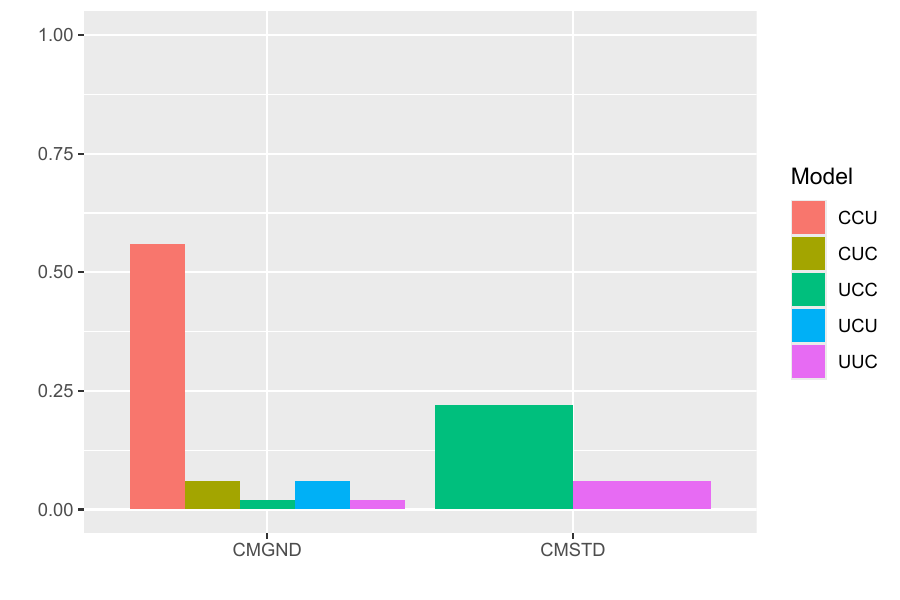}
	\caption{Proportion of times models selected by BIC for both CMGND and CMSTD models.}
	\label{model_prop_fin_app}
\end{figure}

Tables \ref{BICstoxx50e} in Appendix \ref{sec.appendix} report the BIC values for the CMGND, CMND and CMSTD. Mixture models with three components were also considered in the analysis, but were consistently outperformed by two-component models in terms of BIC. The most evident result is that the BIC criterion selects the CMGND model in 72\% of the cases for the 50 constituents of the Euro Stoxx 50 index. For most of the stocks analysed, CMGND models offer a better trade-off between goodness-of-fit and parsimony compared to CMND and CMSTD.\\ 
Figure \ref{model_prop_fin_app} presents the proportion of BIC's selected models for both the CMGND and CMSTD frameworks. For the CMGND specification, the model most frequently selected is the CCU, while for the CMSTD, the UCC model appears most frequently. These findings highlight the ability of CMGND models to adapt to the empirical characteristics of financial returns. It is well known that daily log-returns have a mean close to zero and primarily differ in the behaviour observed in the tails of the distribution. In fact, the CCU model captures this specific feature by imposing a common mean across components, a constraint that neither \textit{mclust} nor \textit{teigen} allow when estimating mixtures of Normal and Student-\textit{t} distributions.

\begin{figure}[H]
\centering
	\includegraphics[width=\textwidth]{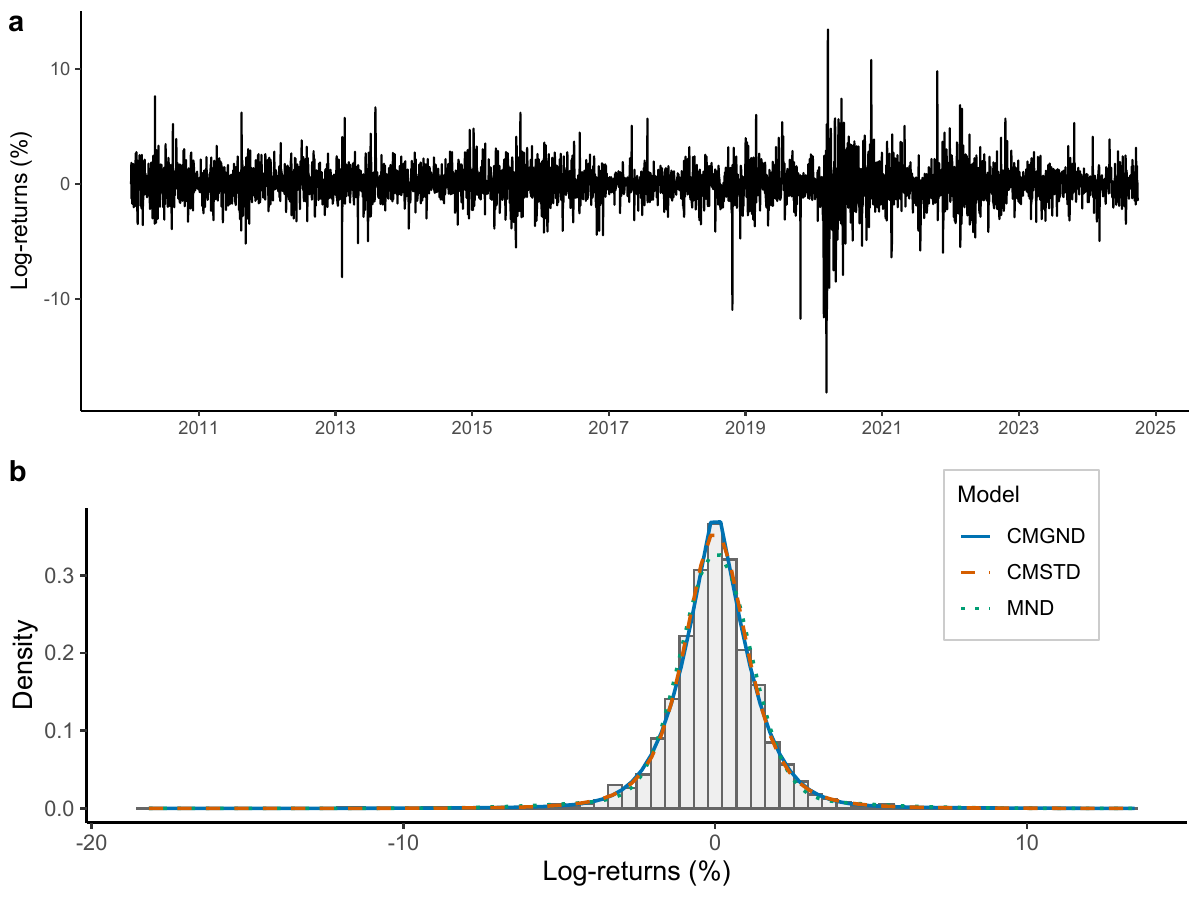}
	\caption{Log-returns (\%) of ABI.BE panel a, and estimated densities panel b.}
	\label{fin_example_den}
\end{figure}

The fact that CMND is never selected by the BIC indicates that a model assuming normal distributions for mixture components is inadequate for describing the complexity of equity returns, which are known to exhibit non-normality such as skewness and high kurtosis. 
This is consistent with the descriptive statistics reported in Table \ref{tab.comp.bs}, which show that all stocks have empirical kurtosis greater than three and the Jarque-Bera (JB) test rejects the normality hypothesis for all stocks. An example is provided for ABI.BE (Anheuser-Busch InBev), the first alphabetically ordered constituent of the SX5E. Panel (a) of Figure \ref{fin_example_den} shows the log-returns (\%) of ABI.BE, highlighting two prominent periods of high volatility: the Covid-19 pandemic (2020-2021) and the onset of the Russia-Ukraine war (February 2022). The daily returns of ABI.BE show clear deviations from normality, as evidenced by the presence of heavy tails, excess kurtosis, and negative skewness (see Table \ref{tab.comp.bs}). Panel (b) presents the estimated return densities for ABI.BE, while the estimated parameters are reported in Table \ref{ABI.BEest}. Among the models considered, the best performing specification is the CCU for the CMGND model, the UCC for the CMSTD model, and the UUU for the CMND model. In particular, the estimated densities produced by the CMGND and CMSTD models are closely aligned; however, the CMGND model better captures the central peak of the distribution. In contrast, the CMND model does not provide an adequate estimate of the central peak.

In addition, the CMGND model provides better financial interpretability by identifying two components with different shape parameters: one representing stable market conditions and the other capturing periods of turmoil \citep{Kon1984,Behr2009,Duttilo2024,Iannone2025}. The stable component has tails that fall between those of the Laplace and normal distributions ($1 < \nu_1 < 2$), while the turmoil component has tails much heavier than the Laplace distribution, as indicated by $\nu_2<1$. In this context, a smaller shape parameter leads to thicker tails and a greater standard deviation, whereas a larger shape parameter results in thinner tails and lower standard deviation. In contrast, the CMSTD model identifies two components based solely on location parameters, a simplistic outcome that lacks a deeper financial understanding. 

The results show that CMGND emerges as a competitive, if not superior, model to CMSTD, which is commonly used for modeling return distributions \citep{Massing2021}. This is a significant outcome, as it suggests that the additional flexibility provided by the generalized normal distribution (which includes the normal and Laplace distributions as special cases and allows modelling of various kurtosis shapes) is advantageous for analysing real financial data.

\begin{table}[H]
\caption{Estimation results for ABI.BE.}
\label{ABI.BEest}
\centering
\begin{tabular}{lccc}
\toprule
&\multicolumn{1}{c}{CMGND}&\multicolumn{1}{c}{CMND}&
\multicolumn{1}{c}{CMSTD}\\
\midrule
$\pi_1$& 0.8625   & 0.8706& 0.5258\\
$\mu_1$&0.0183& 0.0217& 0.0315\\
$\sigma_1$& 1.3525& 1.2189& 1.0765\\
$\nu_1$ & 1.3802&-& 3.3356\\
$\pi_2$ & 0.1375 & 0.1294& 0.4742\\
$\mu_2$ & 0.0183 &-0.0486& 0.0025\\
$\sigma_2$ &1.3525& 12.2199& 1.0765\\
$\nu_2$ &0.7937&-& 3.3356\\
\midrule
$\log L(\widehat{\theta})$& -6716.82& -6755.81&-6717.75\\
BIC & \textbf{13474.83}& 13552.81 & 13476.70\\
\bottomrule
\end{tabular}
\begin{tablenotes}
\centering
\item[]{\footnotesize Note. In bold, the best model according to the BIC.}
\end{tablenotes}
\end{table}

\section{Concluding remarks} \label{Conclusions}
This study addresses the topic of modelling complex distributions through constrained mixtures of generalized normal distributions. Through simulations and analyses of real-world data, the capacity to accurately estimate the parameters of the mixture components and moments of marginal distributions is evaluated, considering varying levels of component overlap and different sample sizes. 

A primary benefit of CMGND models lies in their parametric efficiency. By imposing equality constraints, for location, scale and/or shape parameters, across any subset of mixture components, the number of free parameters is reduced. This parsimony is particularly advantageous when dealing with a large number of components or limited sample sizes, potentially leading to more robust and interpretable models. The use of constraints can highlight similarities between different components of the mixture. For example, a constraint on the shape parameter might suggest that different subgroups within the data share a similar tail behaviour.

The results of numerical experiments indicate that correctly specified constraints can lead to more accurate parameter estimates (lower RMSE) compared to unconstrained MGND models. This is particularly noticeable for parameters that are subject to constraints and in situations characterised by higher overlap between mixture components. Interestingly, improved estimation accuracy can extend to unconstrained parameters within the same components because of the interdependence of parameter estimates.

The use of the BIC criterion for optimal model selection is analysed. Selecting the correct model can be challenging. Model selection criteria like BIC depend on the overlap of components and sample size, with a tendency to favour more parsimonious (constrained) models even when the true model is unconstrained, especially with smaller sample sizes and higher overlap between components. However, even in the presence of model misspecification, the resulting BIC's selected model ensures comparable or even better estimation results with respect to the true model. In particular, imposing constraints on the shape parameter can be particularly effective in capturing the kurtosis of the marginal distribution. 

The results of the real application suggest that CMGND models prove to be a strong candidate for financial data modelling showing strong competitiveness against the mixture of Student's t distributions that currently serve as benchmarks for estimating financial data distribution.

The proposed mixture model can also be extended by introducing the multivariate version \citep{Allili2012} to analyse the correlated data. Another extension could be to develop a time-varying CMGND model based on the score-driven framework of \cite{Harvey2013} where parameters such as the scale or the shape of the distribution are allowed to be dynamic in time.

As a final note, we emphasise that the approach proposed is in fact novel in the literature concerning the implementation of constraints in mixture models. Both the \textit{mclust} package (mixtures of normal distributions) and the \textit{teigen} package (mixtures of student-t distributions) handle constraints across all components of the mixture. They may not support the ability to impose constraints on a specific subset of mixture components. In this respect, our proposal offers more flexibility by allowing constraints (on location, scale, and shape) to be imposed on any subset of the mixture components. As noted above, when the shape parameter $\nu$ is set to 2, the GND becomes equivalent to the normal distribution. This implies that the constrained models proposed with $\nu=2$ have constrained mixtures of normals as a special case, and importantly, they retain the flexibility to apply constraints to subsets of components, which is a limitation of standard packages like \textit{mclust} when dealing with normal mixtures. Another nested model could be obtained by setting $\nu=1$, leading to constrained mixtures of Laplace distributions. Furthermore, none of the current implementations of constrained mixture models allows the possibility of constraining the location parameter. This constraint can be useful in certain application contexts, as we have already observed in the application section.

Therefore, the ability to implement constraints on arbitrary subsets of components, not just across the entire mixture, represents a key novelty and a significant advantage of the constrained mixtures of the generalized normal distribution framework compared to the constraint handling capabilities of existing R packages. This increased flexibility allows for the development of more parsimonious and potentially better fitting models in various applications.

\bibliographystyle{apalike}
\bibliography{References.bib}

\textbf{Disclosure statement} The authors report that there are no competing interests to declare.

\textbf{Data availability statement} 
The authors do not have permission to share the data. To recover the data, please contact the provider LSEG Data \& Analytics.

\appendix
\setcounter{table}{0}
\renewcommand{\thetable}{A\arabic{table}}

\section{Tables}\label{sec.appendix}

\begin{table}[H]
\caption{Descriptive statistics of daily log-returns on SX5E constituents.}
\label{tab.comp.bs}
\centering
\resizebox{12.5cm}{!}{
\begin{tabular}{lccccccccc}
\toprule
\textbf{Tickers} & \textbf{N} & \textbf{Mean} & \textbf{Median} & \textbf{Std} & \textbf{Skewness} & \textbf{Kurtosis} & \textbf{Min} & \textbf{Max} & \textbf{JB} \\ 
  \midrule
ABI.BE    & 3786 & 0.0126  & 0.0155  & 1.6258 & -0.5261 & 15.2961 & -18.1622 & 13.4532 & 24057  \\
AD.NL     & 3786 & 0.0319  & 0.067   & 1.2615 & -0.3433 & 8.6646  & -10.0198 & 7.7183  & 5145   \\
ADS.DE    & 3749 & 0.0483  & 0       & 1.8658 & 0.2118  & 11.8965 & -16.6886 & 19.3787 & 12409  \\
ADYEN.NL  & 1622 & 0.0684  & 0.1629  & 3.1286 & -1.917  & 50.6746 & -49.4034 & 32.0764 & 155007 \\
AI.FR     & 3786 & 0.0361  & 0.0551  & 1.3019 & -0.1667 & 7.7453  & -11.8337 & 8.1161  & 3576   \\
AIR.FR    & 3786 & 0.0587  & 0.0772  & 2.1305 & -0.3696 & 16.6907 & -25.0623 & 18.6175 & 29692  \\
ALV.DE    & 3749 & 0.0326  & 0.069   & 1.5609 & -0.3385 & 14.1051 & -16.6382 & 14.6728 & 19362  \\
ASML.NL   & 3786 & 0.0902  & 0.1274  & 1.9645 & -0.0342 & 6.535   & -13.1142 & 13.0633 & 1976   \\
BAS.DE    & 3749 & 0.0028  & 0.0496  & 1.6847 & -0.2498 & 6.7933  & -12.5494 & 10.1917 & 2291   \\
BAYN.DE   & 3749 & -0.0189 & 0       & 1.7967 & -0.7681 & 11.7311 & -19.798  & 9.8678  & 12294  \\
BBVA.ES   & 3771 & -0.0071 & 0       & 2.1928 & -0.0765 & 10.1088 & -17.649  & 19.9073 & 7956   \\
BMW.DE    & 3749 & 0.0235  & 0.032   & 1.8212 & -0.3276 & 7.8811  & -13.8933 & 13.5163 & 3795   \\
BN.FR     & 3786 & 0.0109  & 0.0175  & 1.237  & -0.1645 & 7.0022  & -8.8931  & 7.4267  & 2549   \\
BNP.FR    & 3786 & 0.0027  & 0.0407  & 2.2237 & -0.1236 & 11.1418 & -19.1166 & 18.9768 & 10482  \\
CS.FR     & 3786 & 0.0192  & 0.0792  & 1.9485 & -0.165  & 14.6239 & -16.8196 & 19.7782 & 21360  \\
DB1.DE    & 3749 & 0.0359  & 0.0504  & 1.5239 & -0.4082 & 9.8567  & -12.5985 & 12.3142 & 7460   \\
DG.FR     & 3786 & 0.0258  & 0.0498  & 1.6607 & -0.3773 & 15.9178 & -18.7227 & 17.2666 & 26448  \\
DHL.DE    & 3749 & 0.0259  & 0.0701  & 1.5913 & -0.2454 & 7.8797  & -12.8063 & 11.7308 & 3764   \\
DTE.DE    & 3749 & 0.0261  & 0.0105  & 1.3685 & -0.4423 & 9.8393  & -11.2673 & 10.6715 & 7440   \\
EL.FR     & 3786 & 0.0432  & 0.0489  & 1.4374 & 0.0345  & 7.8759  & -10.9702 & 11.1999 & 3758   \\
ENEL.IT   & 3748 & 0.0144  & 0.0323  & 1.6522 & -0.9864 & 14.0645 & -22.1228 & 7.6674  & 19753  \\
ENI.IT    & 3748 & -0.0068 & 0.0602  & 1.6936 & -1.1388 & 21.2244 & -23.3851 & 13.9159 & 52743  \\
IBE.ES    & 3771 & 0.0196  & 0.0444  & 1.4705 & -0.3495 & 11.6154 & -15.1554 & 13.3703 & 11756  \\
IFX.DE    & 3749 & 0.0531  & 0.0625  & 2.2842 & -0.2225 & 6.3665  & -17.0262 & 13.0759 & 1805   \\
INGA.NL   & 3786 & 0.0222  & 0.0373  & 2.3247 & -0.1945 & 12.2608 & -21.5324 & 21.9838 & 13572  \\
ISP.IT    & 3748 & 0.0078  & 0.0565  & 2.405  & -0.6557 & 12.0401 & -26.0594 & 17.962  & 13049  \\
ITX.ES    & 3771 & 0.0468  & 0       & 1.6206 & 0.2489  & 7.5087  & -11.1276 & 13.1323 & 3239   \\
KER.FR    & 3786 & 0.0298  & 0.032   & 1.846  & -0.1225 & 7.4353  & -13.1416 & 10.0708 & 3118   \\
MBG.DE    & 3749 & 0.0169  & 0.0391  & 1.9621 & -0.0801 & 15.3704 & -20.8896 & 24.1193 & 23940  \\
MC.FR     & 3786 & 0.0589  & 0.0719  & 1.7051 & 0.1027  & 6.251   & -9.0777  & 12.0552 & 1677   \\
MUV2.DE   & 3749 & 0.0405  & 0.0852  & 1.4875 & -0.2356 & 21.9209 & -19.5309 & 18.3778 & 56027  \\
NDA.FI.FI & 3695 & 0.0118  & 0.0568  & 1.7814 & -0.4705 & 8.756   & -15      & 12.2009 & 5246   \\
NOKIA.FI  & 3695 & -0.0248 & 0.0433  & 2.4486 & -0.7681 & 20.8655 & -26.5878 & 29.2226 & 49565  \\
OR.FR     & 3786 & 0.0423  & 0.0422  & 1.3715 & 0.1117  & 6.1634  & -7.883   & 8.1003  & 1590   \\
PRX.NL    & 1303 & 0.0126  & -0.0197 & 2.5987 & 0.1068  & 10.5524 & -19.015  & 21.4164 & 3113   \\
RACE.IT   & 2235 & 0.1018  & 0.1098  & 1.7443 & -0.0381 & 8.208   & -10.8247 & 10.4542 & 2534   \\
RI.FR     & 3786 & 0.0194  & 0.035   & 1.3185 & -0.203  & 6.744   & -10.3501 & 7.5608  & 2242   \\
RMS.FR    & 3786 & 0.083   & 0.1071  & 1.5862 & -0.0266 & 8.9923  & -12.5081 & 14.0847 & 5674   \\
SAF.FR    & 3786 & 0.0715  & 0.0492  & 1.997  & -0.5218 & 22.6549 & -25.9726 & 19.0064 & 61187  \\
SAN.ES    & 3771 & -0.0223 & 0.0174  & 2.183  & -0.2228 & 12.5898 & -22.1724 & 20.8774 & 14501  \\
SAN.FR    & 3786 & 0.0159  & 0.0386  & 1.4176 & -1.1407 & 18.5638 & -20.9893 & 6.2345  & 39082  \\
SAP.DE    & 3749 & 0.0476  & 0.0783  & 1.4756 & -1.2883 & 28.0599 & -24.7661 & 11.8219 & 99254  \\
SGO.FR    & 3786 & 0.02    & 0.0168  & 1.9277 & -0.4681 & 10.0025 & -18.7602 & 11.2554 & 7885   \\
SIE.DE    & 3749 & 0.0306  & 0.0484  & 1.6214 & -0.1148 & 7.906   & -13.5774 & 10.9387 & 3774   \\
STLAM.IT  & 3748 & 0.0447  & 0.0826  & 2.5407 & -0.4863 & 7.9363  & -19.6792 & 15.1865 & 3960   \\
SU.FR     & 3786 & 0.0468  & 0.0877  & 1.8194 & -0.1478 & 6.9845  & -15.1032 & 11.3445 & 2523   \\
TTE.FR    & 3786 & 0.0085  & 0.0656  & 1.6322 & -0.5328 & 15.685  & -18.1622 & 14.0407 & 25596  \\
UCG.IT    & 3748 & -0.0163 & 0.0424  & 2.835  & -0.3809 & 9.7444  & -27.1658 & 19.0067 & 7205   \\
VOW3.DE   & 3749 & 0.0142  & 0       & 2.14   & -0.501  & 12.7985 & -22.0877 & 17.434  & 15176  \\
WKL.NL    & 3786 & 0.0606  & 0.0849  & 1.2651 & -0.427  & 7.612   & -10.2898 & 7.6426  & 3476  \\
\bottomrule
\end{tabular}
}
\end{table}

\begin{table}[H]
\caption{BIC of fitted mixture models for daily log-returns on SX5E constituents for $K=2$.}
\label{BICstoxx50e}
\centering
\resizebox{6.4cm}{!}{
\begin{tabular}{lccc}
\toprule
Ticker   & CMND      & CMGND   & CMSTD  \\ \midrule
ABI.BE   & 13552.81 & \textbf{13474.83} & 13476.7           \\
AD.NL    & 11876.22 & \textbf{11853.82} & 11858.31          \\
ADS.DE   & 14710.26 & \textbf{14669.49} & 14672.9           \\
ADYEN.NL & 7790.79  & 7740.06           & \textbf{7735.14}  \\
AI.FR    & 12389.09 & 12360.45          & \textbf{12359.97} \\
AIR.FR   & 15635.29 & \textbf{15538.93} & 15542.12          \\
ALV.DE   & 13045.93 & 12974.68          & \textbf{12967.72} \\
ASML.NL  & 15529.75 & \textbf{15495.49} & 15502.42          \\
BAS.DE   & 14177.78 & \textbf{14156.94} & 14159.25          \\
BAYN.DE  & 14505.32 & \textbf{14431.3}  & 14437.82          \\
BBVA.ES  & 16077.94 & \textbf{16016.64} & 16020.48          \\
BMW.DE   & 14687.73 & \textbf{14647.02} & 14647.07          \\
BN.FR    & 11978.5  & 11956.02          & \textbf{11955.16} \\
BNP.FR   & 16069.61 & \textbf{16013.69} & 16019.1           \\
CS.FR    & 14789.35 & 14710.26          & \textbf{14705.13} \\
DB1.DE   & 13310.29 & 13260.27          & \textbf{13258.47} \\
DG.FR    & 13758.11 & \textbf{13681.71} & 13689.43          \\
DHL.DE   & 13694.89 & \textbf{13659.64} & 13662.19          \\
DTE.DE   & 12354.66 & 12307.85          & \textbf{12306.08} \\
EL.FR    & 13031.88 & \textbf{12996.34} & 12998.98          \\
ENEL.IT  & 13982.38 & \textbf{13928.74} & 13930.17          \\
ENI.IT   & 13932.23 & 13865.36          & \textbf{13852.06} \\
IBE.ES   & 12908    & \textbf{12856.41} & 12857.09          \\
IFX.DE   & 16590.84 & \textbf{16550.16} & 16562.18          \\
INGA.NL  & 16296.26 & \textbf{16203.49} & 16209.45          \\
ISP.IT   & 16507.41 & \textbf{16437.95} & 16442.12          \\
ITX.ES   & 14005.65 & \textbf{13975.82} & 13977.09          \\
KER.FR   & 14896.56 & \textbf{14866.75} & 14872.65          \\
MBG.DE   & 15102.14 & \textbf{14991.91} & 14995.08          \\
MC.FR    & 14439.39 & \textbf{14419.57} & 14424.34          \\
MUV2.DE  & 12683.35 & 12613.07          & \textbf{12609.83} \\
NDA.FI   & 14178.05 & \textbf{14158.7}  & 14161.36          \\
NOKIA.FI & 15910.56 & 15825.46          & \textbf{15824.81} \\
OR.FR    & 12821.62 & 12804.67          & \textbf{12804.5}  \\
PRX.NL   & 6028.03  & \textbf{6003.43}  & 6003.5            \\
RACE.IT  & 8511.06  & \textbf{8497.6}   & 8505.43           \\
RI.FR    & 12489.29 & 12473.88          & \textbf{12473.62} \\
RMS.FR   & 13704.87 & \textbf{13666.19} & 13667.33          \\
SAF.FR   & 14964.12 & \textbf{14852.89} & 14855.51          \\
SAN.ES   & 16096.99 & \textbf{16017.34} & 16024.28          \\
SAN.FR   & 12936    & 12898.92          & \textbf{12886.99} \\
SAP.DE   & 12932.59 & 12864.81          & \textbf{12857.47} \\
SGO.FR   & 15241.26 & \textbf{15173.78} & 15180.52          \\
SIE.DE   & 13848.13 & \textbf{13822.72} & 13825.73          \\
STLAM.IT & 17201.79 & \textbf{17145.92} & 17155.89          \\
SU.FR    & 14938.45 & \textbf{14883.42} & 14887.73          \\
TTE.FR   & 13830.43 & \textbf{13743.44} & 13750.41          \\
UCG.IT   & 17852.52 & \textbf{17800.29} & 17804.93          \\
VOW3.DE  & 15782.99 & \textbf{15686.4}  & 15695.19          \\
WKL.NL   & 12121.32 & \textbf{12105.46} & 12106.23          \\
\bottomrule
\end{tabular}
}
\begin{tablenotes}
\centering
\item[]{\footnotesize Note. In bold, the best model according to the BIC.}
\end{tablenotes}
\end{table}

\end{document}